\documentclass[aps,pra,twocolumn,showpacs]{revtex4}
\usepackage{amsmath,amssymb,graphicx,bbold}
\usepackage{graphicx}  % needed for figures
\usepackage{dcolumn}   % needed for some tables
\usepackage{bm}        % for math
\usepackage{color}

\providecommand{\abs}[1]{\left\lvert#1\right\rvert}

\providecommand{\bra}[1]{\langle #1 \rvert}
\providecommand{\ket}[1]{\lvert #1 \rangle}
\providecommand{\braket}[2]{\langle #1 \rvert #2 \rangle}

\begin{document}

 \title{Topological phase structure of entangled qudits}
\author{A. Z. Khoury and L. E. Oxman}

\affiliation{
Instituto de F\'\i sica, Universidade Federal Fluminense,
24210-346 Niter\'oi - RJ, Brasil}
\date{\today}

\begin{abstract}
We discuss the appearance of fractional topological phases on cyclic evolutions 
of entangled qudits. The original result reported in Phys. Rev. Lett. \textbf{106}, 
240503 (2011) is detailed and extended to qudits of different dimensions. The 
topological nature of the phase evolution and its restriction to fractional values 
are related to both the structure of the projective space of states and entanglement. 
For maximally entangled states of qudits with the same Hilbert space dimension, 
the fractional geometric phases are the only ones attainable under local SU($d$) 
operations, an effect that can be experimentally observed through conditional 
interference. 
\end{abstract}
%\vskip2pc 

\maketitle

\section{introduction}
\label{intro}

Geometrical phases are a remarkable property of quantum phase 
evolutions, related to holonomies in the parameter space 
characterizing the quantum state vectors. The standard 
example is the phase acquired by a spin $1/2$ particle 
undergoing a cyclic evolution described by a closed path 
in the Bloch sphere. The role of holonomies in the quantum 
phase evolution was pointed out by M. V. Berry \cite{berry} 
in connection with adiabatic transformations driven by a 
slowly varying time dependent Hamiltonian. An analogous 
effect was studied in a seminal work by S. Pancharatnam 
\cite{pancha} in a more elementary system, the polarization 
transformations in classical wave optics. A beautiful 
generalization of the Pancharatnam results to paraxial 
mode transformations was theoretically 
proposed in Refs. \cite{vanenk,padgett} and experimentally 
demonstrated in Ref. \cite{galvez} where a Poincar\'e sphere 
representation was used for first order paraxial modes. 
This representation was also used to discuss the geometric 
phase conjugation in an optical parametric oscillator \cite{geophaseopo}. 
More recently, the geometric representation of higher order paraxial 
modes have been discussed in Refs. \cite{milione1,milione2}. 
Another fundamental contribution to the theory of geometric phases 
was given by Mukunda and Simon in Refs. \cite{smukunda,smukunda2}, 
where the kinematical aspects of the quantum state evolution 
were investigated and the geometric phase generalized to non 
adiabatic evolutions.  
Since these seminal contributions, numerous works have been 
devoted to both fundamental and applied aspects of geometric 
phases.

In quantum information science, geometric phases were conceived as a robust 
means for implementing unitary gates useful for quantum computation 
\cite{vedral,zoller}. The role of entanglement in the phase evolution of 
two-qubit systems was investigated in refs.\cite{sjoqvist,sjoqvist2}, and 
the topological nature of the corresponding geometric phases was investigated 
both theoretical \cite{remy,remy2,milman} and experimentally in the context of 
spin-orbit transformations on a paraxial laser beam \cite{topoluff} and in 
nuclear magnetic resonance \cite{nmr}. 
In a recent work, we investigated 
the crucial role played by the dimension of the Hilbert space 
on the topological phases acquired by entangled qudits 
\cite{fracuff,fracuffufmg}. The appearance of fractional phases is a remarkable property 
of two-qudit systems, also shared by multiple qubits \cite{multiqubit,multiqubit2}. 
Multidimensional entangled states can 
be realized on qudits encoded on the transverse position of quantum correlated photon 
pairs generated by spontaneous parametric down conversion 
\cite{sebastiao,sebastiao2,sebastiao3,posinterf,imagepol}.
Fractional phases were originally investigated in quantum Hall systems in 
connection with different homotopy classes in the configuration space of anyons. 
The topological structure of quantum Hall systems has been conjectured  
to be a possible resource for fault tolerant quantum computation \cite{faulttolerant}. 
These potential applications of geometric phases in quantum information science 
motivated a number of articles devoted to their implementation in quantum optical 
systems and their behavior under the influence of different kinds of reservoir 
\cite{mtt1,mtt2,santos1,santos2,santos3,santos4,santos5,santos6}. Decoherence 
is recognized as the main difficulty for quantum information protocols in 
realistic  physical systems. In this sense quantum gates based on 
geometric phases are supposed to be a powerful tool. 
The geometric phase acquired by an open system was 
studied in Ref. \cite{santos3}, where it was shown that the geometric phase 
was insensitive to dephasing processes. 

In the present work we study the geometric phase acquired by entangled 
qudits under local unitary evolutions. The fractional phases predicted 
in Ref. \cite{fracuff} are developed in detail and generalized to qudits 
of different dimensions. A general expression is derived for the two-qudit 
geometric phase in terms of entanglement and the dimensions of their 
Hilbert spaces. 
We also discuss the holonomy of the phase evolution in terms of the 
parameters used to define the local SU($d$) transformations applied to 
each qudit. 
The manuscript is organized as follows, in section \ref{geosingle} 
we discuss the role played by the purity of a single qudit state in the 
geometric phase. 
Since, for pure states, two-qudit entanglement can be quantified by the 
purity of the partial density matrix of each qudit, the results of section 
\ref{geosingle} are used in section \ref{twoqudit} to establish the role of 
entanglement in the geometric phase acquired by a two-qudit state under 
local unitary transformations. In section \ref{examples} we present some 
numerical examples that illustrate the fractional phase values expected 
and the role played by entanglement. Finally, in section \ref{conclusion} 
we summarize our results and briefly discuss some future perspectives.

%##########################################################################################
\section{Topological phases on single qudits}
\label{geosingle}

Initially, we shall examine the properties of the geometric phases on unitary 
evolutions of single qudits and the role of the quantum state purity 
in the geometric phase. Our conclusions will be useful 
since two-qudit entanglement is frequently quantified through the 
purity of the partial density matrix describing one of the qudits. 
Therefore, we start by considering a single qudit initially prepared 
in a quantum state described by a general density matrix 
\begin{equation}
\rho_0=\frac{\mathbb{1}}{d}+q\,\sqrt{\frac{d-1}{d}}\,\mathbf{\hat{q}}\cdot\mathbf{T}\;,
\label{singlerho}
\end{equation}
where 
$\{T_{\alpha}\}$ are $d^2-1$ Hermitian generators of SU($d$) normalized according 
to $\mathrm{Tr}[T_{\alpha}\,T_{\beta}]=\delta_{\alpha\,\beta}\,$. We shall term 
$\mathbf{q}=q\,\mathbf{\hat{q}}\in \mathbb{R}^{d^2-1}\,$ as the \textit{purity vector} 
(for qubits it is the well known Bloch vector), since its absolute 
value is related to the purity of the qudit state: 
$\mathrm{Tr}[\rho_0^2]=q^2+(1-q^2)/d\,$, with $0\leq q \leq 1\,$. 
From the kinematic approach by Mukunda and Simon \cite{smukunda,smukunda2}, 
the geometric phase acquired by a time evolving pure state $\ket{\psi(t)}$ is given by 
\begin{eqnarray}
\phi_g &=& \arg\langle\psi(0)|\psi(t)\rangle + 
i\int dt \,\,\langle\psi(t)|\dot{\psi}(t)\rangle
\nonumber\\
&=& \arg\{\mathrm{Tr}[\rho_0\,U(t)]\}+i\int dt\,\, \mathrm{Tr}[\rho_0\,U^{\dagger}\dot{U}]
 \;. 
\label{phigsingle}
\end{eqnarray}
where $\rho_0=\ket{\psi(0)}\bra{\psi(0)}\,$ is the density matrix of the initial state 
evolving under the action of the unitary operator $U(t)\,$. 
The second equality 
allows for a natural extension of the geometric phase for mixed states by taking the 
general density matrix (\ref{singlerho}). 
Also, it is important to rule out from the geometric 
phase any explicit phase evolution contained in $U(t)$. Let 
$U(t)=e^{i\phi(t)}\,\bar{U}(t)\,$, where $\bar{U}(t)\in SU(d)\,$ for all $t$, with initial conditions 
$\phi(0)=0$ and $\bar{U}(0)=\mathbb{1}\,$. It is straightforward 
to show that the explicit phase $\phi(t)$ does not contribute to the geometric phase, 
which is then given in terms of the SU($d$) sector only:
\begin{eqnarray}
\phi_g &=& \bar{\phi}_{\,tot}+i\int dt\,\, \mathrm{Tr}[\rho_0\,\bar{U}^{\dagger}\dot{\bar{U}}]
 \;, 
\label{phigsingle2}
\end{eqnarray}
where we defined 
\begin{eqnarray}
\bar{\phi}_{\,tot}\equiv\arg\{\mathrm{Tr}[\rho_0\,\bar{U}(t)]\}\;. 
\end{eqnarray}
Now it is useful to recall that for a general invertible matrix $\mathbb{A}$ we have \cite{cookbook}:
\begin{equation}
\frac{d\left(\det\mathbb{A}\right)}{dt}=
\left(\det\mathbb{A}\right)\,\mathrm{Tr}\left[\mathbb{A^{-1}}\frac{d\mathbb{A}}{dt}\right]\;.
\label{useful0}
\end{equation}
Since the evolution $\bar{U}(t)$ is closed in SU($d$), we readily deduce 
that $\mathrm{Tr}[\bar{U}^{\dagger}\dot{\bar{U}}]=0\,$, 
so that $\bar{U}^{\dagger}\dot{\bar{U}}$ can be written as a linear combination 
of the SU($d$) generators. Moreover, this linear 
combination must only involve purely imaginary coefficients, since 
$d(\bar{U}^{\dagger}\,\bar{U})/dt=0\Rightarrow\bar{U}^{\dagger}\dot{\bar{U}}=
-(\bar{U}^{\dagger}\dot{\bar{U}})^{\dagger}\,$. 
Thus, we can introduce a useful \textit{velocity} vector 
$\mathbf{u}\in\mathbb{R}^{d^2-1}$ such that 
$\bar{U}^{\dagger}\dot{\bar{U}}=i\,\mathbf{u}\cdot\mathbf{T}\,$. 
The geometric phase can be expressed in terms of the purity and velocity 
vectors as 
\begin{eqnarray} 
\phi_g &=& \bar{\phi}_{\,tot} - q\,\sqrt{\frac{d-1}{d}}\,\int \,\, \mathbf{\hat{q}}\cdot\mathbf{dx} 
\;, 
\label{phigsingle3} 
\end{eqnarray} 
where $\mathbf{dx}\equiv\mathbf{u}\,dt\,$ is a connection. 

For a pure state, an evolution over a time interval $T$ is considered to be cyclic 
when it takes the system from a given initial state to a physically equivalent final 
state, \textit{i.e.}, when $\braket{\psi(0)}{\psi(T)}=e^{\,i\phi_{\,tot}(T)}\,$. 
This condition can be generalized for mixed states as 
$\mathrm{Tr}[\,\rho_0\,U(T)\,]=e^{i\phi_{\,tot}(T)}\,$. Now, let us inspect 
carefully this condition over totally mixed states: $q=0\,$. 
In this case, it reduces to 
$\mathrm{Tr}[\,U(T)\,]=d\,e^{\,i\phi_{\,tot}(T)}\,$, which implies 
$U(T)=e^{\,i\phi_{\,tot}(T)}\mathbb{1}\,$, and hence 
$\bar{U}(T)=e^{\,i\bar{\phi}_{\,tot}(T)}\mathbb{1}\,$. Since $\bar{U}\in SU(d)\,$, 
$\det\bar{U}=e^{\,id\,\bar{\phi}_{\,tot}(T)}=1\,$, so that $\bar{\phi}_{\,tot}(T)=2\,n\,\pi/d\,$ 
($n\in\mathbb{Z}$). For qubits, this corresponds to the two possible values $0$ or $\pi$. 

Therefore, expression (\ref{phigsingle3}) for a 
completely mixed state reduces to
\begin{equation}
\phi_g=\bar{\phi}_{tot}=\frac{2\,n\,\pi}{d}\;\;(n\in\mathbb{Z})\;.
\label{ff}
\end{equation}
In principle, this result is of little physical relevance, since no interference 
can be measured on completely incoherent states. However, we can anticipate 
its important role on entangled states. Indeed, the partial trace of a maximally 
entangled pure state of a bipartite system produces completely mixed density 
matrices. In this case, we shall see that for cyclic evolutions, driven by local 
unitary operations, only the fractional phases in Eq. (\ref{ff}) can arise. 
However, they can now be measured through conditional interference, as long as 
the overall bipartite state is coherent. We shall put these 
arguments on a more formal ground in section \ref{twoqudit}.

%%%%%%%%%%%%%%%%%%%%%%%%%%%%%%%%%%%%%%%%%%%%%%%%%%%%%%%%%%%%%%%%%%%%%%%%%%%%%%%%%%%%%%%%%%%%%%%

\subsection{The Cartan Sector}
\label{Cartan}

An interesting refinement of the geometric phase structure is obtained 
by identifying the Cartan subalgebra of SU($d$) in the density matrix and the 
evolution operator. In this manner, we will be able to isolate the 
nonholonomic contribution to the geometric phase. 

The first $d-1$ generators $T_1, \dots, T_{d-1}$ can be taken as the diagonal
generators that form the Cartan subalgebra of SU($d$) (the only element for 
SU($2$) is $\sigma_z$), they will 
also be named as $\{H_{\beta}\}$, $\beta = 1,\dots, d-1$.
The set of SU($d$) generators can then be separated as 
$\{T_{\alpha}\}=\{H_{\beta}\}\cup\{P_{\gamma}\}\,$, where  
$\{P_{\gamma}\}$, $\gamma = 1,\dots, d^2-d$, represents the 
remaining $d^2-d$ nondiagonal generators. As a convention, dot 
products involving $\mathbf{H}$ ($\mathbf{P}$) 
will be used to represent Lie algebra elements restricted to the diagonal 
(off-diagonal) sector. 
That is, dot products between the full set of generators $\mathbf{T}$ and 
vectors having the last $d^2-d$ (the first $d-1$) components vanishing. 

Choosing a Hilbert space basis that renders the initial density matrix 
diagonal, we may write 
\begin{eqnarray}
\rho_{0}&=&\frac{\mathbb{1}}{d}+q\,\sqrt{\frac{d-1}{d}}\,\mathbf{\hat{q}}\cdot\mathbf{H} 
\nonumber\\
&=&\frac{\mathbb{1}}{d}+q\,\sqrt{\frac{d-1}{d}}\,\mathrm{diag}\left[x_0\ldots x_{d-1}\right]\;,
\label{singlerhocartan}
\end{eqnarray}
where we defined $x_{\,n}\equiv\bra{n}\,\mathbf{\hat{q}}\cdot\mathbf{H}\,\ket{n}\,$, 
with the properties $\sum_{\,n} x_{\,n}=0$ and $\sum_{\,n} x^2_{\,n}=1\,$. 
In addition, using the coset factorization of the SU($d$) group \cite{coset}, we can write 
\begin{eqnarray}
\bar{U} &=& 
\bar{V}\,\exp\left(i\,\mathbf{h}\cdot \mathbf{H}\right)\;,
\label{UVbarqudit}
\end{eqnarray}
where the parameters $\mathbf{h}$ map an $\mathbb{R}^{d-1}\,$ subspace and 
$\bar{V}\in {\rm SU}(d)/{\rm U}(1)^{d-1}$. The latter manifold can in turn 
be written as a tensor product of different coset spaces \cite{coset}. 
For $SU(2)$ there is only one factor, ${\rm SU}(2)/{\rm U}(1)$, 
which is topologically equivalent to the two-sphere $S^2$. In general, the 
$\bar{V}$ factor can be defined by the following requirement: if 
\begin{equation}
\bar{V} H_{\beta} \bar{V}^{-1} = H_{\beta}\;,
\label{req}
\end{equation}
for every diagonal generator $H_{\beta}$, then necessarily $\bar{V}=1$.

Using this factorization, we have,
\begin{eqnarray}
\bar{U}^{\dagger}\,\dot{\bar{U}}&=&e^{-i\mathbf{h}\cdot\mathbf{H}}\, 
\bar{V}^{\dagger}\,\dot{\bar{V}}\, e^{i\mathbf{h}\cdot\mathbf{H}} 
+ i\,\dot{\mathbf{h}}\cdot\mathbf{H}\;
\label{udaggerudotqudit}
\end{eqnarray}
where the velocity vector associated with the $\bar{V}$-sector can be separated into two orthogonal 
terms $\mathbf{v_{\|}}$ and $\mathbf{v_{\bot}}$, related to the Cartan subalgebra and the 
nondiagonal generators, respectively. They are defined 
according to 
\begin{equation}
\bar{V}^{\dagger}\dot{\bar{V}}=i\,\mathbf{v_{\bot}}\cdot\mathbf{P} + i\,\mathbf{v_{\|}}\cdot\mathbf{H}\,. 
\end{equation}
Now, using the Baker-Campbell-Hausdorff formula and the fact that 
$[H_{\alpha},P_{\beta}]\propto P_{\gamma}\,$, 
it is easy to show that the transformation 
$e^{-i\,\mathbf{h}\cdot\mathbf{H}}\, \bar{V}^{\dagger}\,\dot{\bar{V}}\, 
e^{i\,\mathbf{h}\cdot\mathbf{H}}$ leaves $\mathbf{v_{\|}}\,$ unchanged 
and makes $\mathbf{v_{\bot}}\rightarrow\mathbf{v_{\bot}^{\,\prime}}\,$, 
so that 
\begin{equation}
e^{-i\,\mathbf{h}\cdot\mathbf{H}}\, \bar{V}^{\dagger}\,\dot{\bar{V}}\, 
e^{i\,\mathbf{h}\cdot\mathbf{H}}=i\,\mathbf{v_{\bot}^{\,\prime}}\cdot\mathbf{P} + 
i\,\mathbf{v_{\|}}\cdot\mathbf{H}\;. 
\label{cartantransf}
\end{equation}
Moreover, the orthonormality condition for the generators leads to, 
$\mathrm{Tr}\left[\left(\bar{V}^{\dagger}\,\dot{\bar{V}}\right)^2\right]=
\left|\mathbf{v_{\bot}}\right|^2+\left|\mathbf{v_{\|}}\right|^2$ and 
$\left|\mathbf{v_{\bot}^{\,\prime}}\right|^2=\left|\mathbf{v_{\bot}}\right|^2\,$. 
Therefore, $\mathbf{v_{\bot}^{\,\prime}}$ corresponds to a rotation of $\mathbf{v_{\bot}}$ 
in a subspace orthogonal to the Cartan subspace where both $\mathbf{\hat{q}}$ and 
$\mathbf{v_{\|}}\,$ 
lie ($\mathbf{v_{\bot}^{\,\prime}}\cdot\mathbf{v_{\|}}=
\mathbf{v_{\bot}^{\,\prime}}\cdot\mathbf{\hat{q}}=0$). 
Finally, from Eq.(\ref{cartantransf}) we get 
\begin{eqnarray}
\bar{U}^{\dagger}\,\dot{\bar{U}} &=& 
i\,\mathbf{v_{\bot}^{\,\prime}}\cdot\mathbf{P}+i\,(\mathbf{v_{\|}}+\dot{\mathbf{h}})\cdot\mathbf{H}\;, 
\label{UdaggerUdotqudit}
\end{eqnarray}
which corresponds to the following decomposition of the velocity vector
\begin{eqnarray}
\mathbf{u} &=& 
\mathbf{v_{\bot}^{\,\prime}} + \mathbf{v_{\|}} + \dot{\mathbf{h}}\;. 
\label{decomp-u}
\end{eqnarray}

Since a diagonal representation has been assumed for $\rho_0\,$, only $\mathbf{v_{\|}}$ and
$\dot{\mathbf{h}}$ will contribute to the integral term in the geometric phase. Noting that
\begin{eqnarray}
\mathrm{Tr}\left[\rho_0\,\bar{U}^{\dagger}\,\dot{\bar{U}}\right] &=& 
i\,q\,\sqrt{\frac{d-1}{d}}\,\,\mathbf{\hat{q}}\cdot(\mathbf{v_{\|}}+\mathbf{\dot{h}})\;,
\label{TrUdaggerUdotqudit}
\end{eqnarray}
and replacing in Eq.(\ref{phigsingle3}), we get
\begin{eqnarray} 
\phi_g &=& \bar{\phi}_{\,tot} - q\,\sqrt{\frac{d-1}{d}}\,\left( \mathbf{\hat{q}}\cdot\mathbf{h}(t) + 
\int \,\, \mathbf{\hat{q}}\cdot\mathbf{dx_{\|}}\right)\,, 
\label{phigsinglequdit} 
\end{eqnarray} 
where $\mathbf{dx_{\|}}\equiv\mathbf{v_{\|}}\,dt$ and $\mathbf{h}(0)=\mathbf{0}\,$. 
The integral term represents a path dependent (nonholonomic) contribution, built 
along the path followed on ${\rm SU}(d)/{\rm U}(1)^{d-1}$. 
When cyclic evolutions are considered, this term generalizes to SU($d$) the usual solid 
angle contribution for paths on the Bloch sphere ${\rm SU}(2)/{\rm U}(1)$, obtained for SU($2$). 
Then, we shall define 
\begin{eqnarray} 
\Phi &=& \oint \,\, \mathbf{\hat{q}}\cdot\mathbf{dx_{\|}}\;. 
\label{Omegap} 
\end{eqnarray} 
Let us denote as \textit{partially cyclic} those evolutions that, at a 
given time $\bar{t}$, close a path in the ${\rm SU}(d)/{\rm U}(1)^{d-1}$ sector.
With regard to the total phase, as $\bar{V}(\bar{t})\in {\rm SU}(d)$ and 
$\bar{V}(0)=1$, this would mean that $\bar{V}(\bar{t})$ 
must be the identity matrix times the exponential of a fractional phase. 
However, such $\bar{V}(\bar{t})$ would satisfy the condition (\ref{req}) 
and, as a consequence, it must necessarily be the identity matrix. 
Therefore, for a partially cyclic evolution,
\begin{eqnarray}
\bar{U}(\bar{t}) &=& 
\exp\left(i\,\mathbf{h}(\bar{t})\cdot \mathbf{H}\right)\;,
\label{Ub}
\end{eqnarray}
and the geometric phase is given by,
\begin{eqnarray} 
\phi_g &=& \bar{\phi}_{\,tot} - q\,\sqrt{\frac{d-1}{d}}\,
\left[\mathbf{\hat{q}}\cdot\mathbf{h}(\bar{t}) + \Phi\right]\;. 
\label{phigsinglequditpartial} 
\end{eqnarray} 
\begin{eqnarray}
\bar{\phi}_{\,tot}\equiv\arg \left\{\mathrm{Tr}
\left[\left( \frac{\mathbb{1}}{d}+q\,\sqrt{\frac{d-1}{d}}\,
\mathbf{\hat{q}}\cdot\mathbf{H} \right)\, 
e^{i\,\mathbf{h}(\bar{t})\cdot \mathbf{H}}\right]\right\}\;. 
\nonumber \\ 
\end{eqnarray}

For qubits, the Cartan sector reduces to a single parameter. 
The identification 
of the Cartan sector will be particularly useful to demonstrate the fractional phases 
for dimensions 
$d>2$, since the number of parameters in the nondiagonal sector scales as $d^{\,2}\,$, 
while in the Cartan sector it scales as $d\,$. We will next build a useful representation 
for the Cartan sector that simplifies its parametrization and will be particularly 
useful for experimental proposals. 

Let us now study how the fractional phases, generated in cyclic evolutions, are built. 
To simplify the discussion,
consider evolutions restricted to the Cartan sector
\begin{eqnarray}
\bar{U}(t)=e^{i\,\mathbf{h}(t)\cdot\mathbf{H}}=\mathrm{diag}
\left[e^{i\,\chi_0},\ldots,e^{i\,\chi_{d-1}} \right]\;,
\end{eqnarray}
with $\chi_n(t)\equiv\bra{n}\,\mathbf{h\cdot H}\,\ket{n}$ and 
$\sum_n \chi_n=0\,$. 
For the initial density matrix given by Eq.(\ref{singlerhocartan}), the geometric 
phase can be easily computed
\begin{eqnarray}
\phi_g &=& \bar{\phi}_{tot}-q\,\sqrt{\frac{d-1}{d}}\,\sum_{n=0}^{d-1}\,
x_{\,n}\,\chi_n\;,
\label{phigsinglediagonal}
\end{eqnarray}
where the nontrivial total phase is 
\begin{eqnarray}
\bar{\phi}_{tot}=\arg\left\{\sum_{n=0}^{d-1}\,
\left(\frac{1}{d}+q\,\sqrt{\frac{d-1}{d}}\,x_{\,n}\right)
\,e^{i\,\chi_{\,n}}\right\}\;. 
\label{phitotsinglediagonal}
\end{eqnarray}
Now, for completely mixed states ($q=0$), 
a quite subtle feature of the Cartan sector comes into play. The diagonal 
elements in $\bar{U}(t)$ are phasors in the complex plane. The state evolution will 
be cyclic when these phasors line up, making $\bar{U}$ proportional to the identity 
matrix. This will happen when $\Delta \chi_n\equiv \chi_0-\chi_n=2\,l_n\pi\,$, with 
$l_n\in\mathbb{Z}\,$. However, this alignment can only occur at fractional phase 
values. In order to see this, let us sum up all phase differences and make 
$\sum_{n}\Delta \chi_n=2\pi\, L$, with $L\equiv \sum_n l_n\,$. 
On the other hand, 
\begin{eqnarray}
\sum_{n=0}^{d-1} \Delta \chi_n &=& 
\sum_{n=0}^{d-1} \chi_0 - \sum_{n=0}^{d-1} \chi_n = d\,\chi_0
%\label{Unn}
\end{eqnarray}
what brings us to the fractional solutions $\chi_0=2\pi\, L/d$ and $\chi_n=2\pi\, L/d - 2\pi\,l_n$, as 
expected. Then, the nontrivial total phase is
\begin{eqnarray}
\bar{\phi}_{\,tot} = \frac{2\pi\,L}{d}\;,
\label{Lpisingle}
\end{eqnarray}
and this is the only contribution to the geometric phase acquired by completely mixed 
states.

%%%%%%%%%%%%%%%%%%%%%%%%%%%%%%%%%%%%%%%%%%%%%%%%%%%%%%%%%%%%%%%%%%%%%%%%%%%%%%%%%%%%%%%%%%%%%%%

\subsection{Qubits}
\label{geosinglequbits}

As an example, let us apply the ideas above to the simplest case 
of a single qubit. 
The normalized SU($2$) generators can be written in terms of the 
Pauli matrices; the nondiagonal sector is composed 
by $P_1=\sigma_x/\sqrt{2}$ and $P_2=\sigma_y/\sqrt{2}\,$, while the 
Cartan sector corresponds to
$H_1=\sigma_z/\sqrt{2}\,$. 
Let us use a basis such that the initial density matrix is diagonal:
\begin{eqnarray}
\rho_0 &=& 
\left[
\begin{matrix}
\frac{1+q}{2} & 0\\
\\
0 & \frac{1-q}{2}
\end{matrix}
\right]= \frac{\mathbb{1}+q\,\sigma_z}{2}\;,
\label{rho0qubit}
\end{eqnarray}
where $\{\ket{0},\ket{1}\}$ are the eigenvectors of $\sigma_z$ with eigenvalues $\{+1,-1\}$, 
respectively, and $0\leq q\leq 1\,$. This initial state corresponds to the purity vector 
\begin{equation}
\mathbf{q}=(q,0,0)\;.
\label{q0qubit} 
\end{equation}
Then, suppose this qubit evolves under the action of a general SU($2$) matrix 
\begin{eqnarray}
\bar{U}(\theta,\varphi,\chi) &=& 
\bar{V}(\theta,\varphi)\,e^{i\chi\,\sigma_z}\;,
\nonumber\\
\bar{V}(\theta,\varphi) &=& 
\exp\left(i\,\theta\,\mathbf{\hat{p}}\cdot\mathbf{P}\right)
\nonumber\\
&=&\left[
\begin{matrix}
\cos\frac{\theta}{2} & i\,\sin\frac{\theta}{2}\,e^{-i\varphi}\\
i\,\sin\frac{\theta}{2}\,e^{i\varphi} & \cos\frac{\theta}{2}
\end{matrix}
\right]\;,
\label{UVbarqubit}
\end{eqnarray}
where $\mathbf{\hat{p}}=(0, \cos\varphi , \sin\varphi)\,$, 
$\mathbf{h}=\sqrt{2}\,(\chi, 0,0)\,$, and 
$\varphi(t)$, $\theta(t)$, and $\chi(t)$ are time dependent real parameters with 
initial conditions $\varphi(0)=\theta(0)=\chi(0)=0\,$. Here, $\varphi(t)$ and $\theta(t)$ 
can be identified with the angular coordinates on the Bloch sphere representation of 
a pure state. In fact, they are precisely the coordinates of the evolving state when 
it is initially prepared in $\ket{0}$ ($q=1$). 
Therefore, we identify the state evolution as an explicit phase evolution 
$\chi(t)$ (not to be confused with the explicit phase $\phi(t)$ discarded above, since 
$\bar{U}$ is already an SU($2$) matrix) and a path $(\theta(t),\varphi(t))$ on the Bloch sphere. 

The velocity vector $\mathbf{u}\in\mathbb{R}^3$ can be computed from the decomposition of 
$\bar{U}^{\dagger}\,\dot{\bar{U}}$ in terms of the SU($2$) generators (Pauli matrices). 
It is more elegant to do it in two steps. Initially, we note that 
\begin{equation}
\bar{U}^{\dagger}\,\dot{\bar{U}}=
e^{-i\chi \sigma_z}\,\bar{V}^{\dagger}\,\dot{\bar{V}}\,e^{i\chi\sigma_z}
+i\,\chi\,\sigma_z\;, 
\label{u+udot}
\end{equation}
and write $\bar{V}^{\dagger}\,\dot{\bar{V}}=i\,\mathbf{v}\cdot\mathbf{T}\,$, where $\mathbf{v}$ 
is the velocity vector along the path followed on the Bloch sphere. 
From eq.(\ref{UVbarqubit}) we obtain ${\rm\bf v} = (v_h,v_{p1},v_{p\,2})$, where
\begin{eqnarray}
v_{h} &=&  \sqrt{2}\,\dot{\varphi}\,\sin^2\left(\frac{\theta}{2}\right)
\nonumber \\
v_{p1} &=& \frac{1}{\sqrt{2}}\left(\dot{\theta}\,\cos\varphi - 
\dot{\varphi}\,\sin\theta\,\sin\varphi\right) 
\nonumber\\
v_{p\,2} &=& \frac{1}{\sqrt{2}}\left(\dot{\theta}\,\sin\varphi + 
\dot{\varphi}\,\sin\theta\,\cos\varphi\right)\;.
\end{eqnarray}
The first term in eq.(\ref{u+udot}) amounts to a rotation 
of $\mathbf{v}$ by an angle $2\chi$, generated by $\sigma_z$, so that 
\begin{eqnarray}
\mathbf{u}&=&\mathbf{v^{\,\prime}} + \sqrt{2}\,(\dot{\chi},0,0)\;, 
\end{eqnarray}
where $\mathbf{v^{\,\prime}}=R_z(2\chi)\,\mathbf{v}\,$. 
This rotation leaves the Cartan component of $\mathbf{v}$ unchanged
so that, in this 
SU($2$) parametrization, the connection becomes 
\begin{eqnarray}
\mathbf{\hat{q}}\cdot\mathbf{dx}=\sqrt{2}\left(\,d\chi + \sin^2(\theta/2)\,d\varphi\right)\;.
\label{intusingle}
\end{eqnarray}
The first term is the holonomic contribution, while the second one (nonholonomic) is built 
along the path followed on the Bloch sphere. For a closed path, it gives 
\begin{eqnarray}
\Phi=\sqrt{2}\oint  \sin^2\frac{\theta}{2}\;d\varphi = 
\frac{1}{\sqrt{2}}\int\int\,\sin\theta\;d\theta\,d\varphi = \frac{\Omega}{\sqrt{2}}\;,
\nonumber\\
\label{phigqubit2}
\end{eqnarray}
where the second equality results from Green's Theorem, giving the usual 
solid angle contribution $\Omega$, enclosed on the Bloch sphere. 

Now, let us inspect these contributions for partially cyclic evolutions, that is when 
$(\theta(t),\varphi(t))$ follows a closed path on the 
Bloch sphere over a time interval $\bar{t}$, but $\chi(t)$ does not complete a full cycle. 
In this case, $\bar{U}(\bar{t})=e^{i\,\chi(\bar{t})\sigma_z}$ and 
\begin{equation} 
\bar{\phi}_{\,tot} = \arg\left\{\cos\chi + i\,q\,\sin\chi\right\}\;, 
\label{trace}
\end{equation} 
which gives
\begin{eqnarray}
\phi_g = \arctan\left(q\,\tan\chi\right) 
- q\,\left(\chi + \frac{\Omega}{2}\right)\;.
\label{phigqubit3}
\end{eqnarray}
For an initial pure state $\ket{0}$ ($q=1$), one obtains the usual 
solid angle expression $\phi_g=-\Omega/2\,$. For completely mixed states 
($q=0$) the integral terms vanish and the only possible geometric 
phases are $0$ or $\pi\,$. 

%%%%%%%%%%%%%%%%%%%%%%%%%%%%%%%%%%%%%%%%%%%%%%%%%%%%%%%%%%%%%%%%%%%%%%%%%%%%%%%%%%%%%%%%%%%%%%%

\subsection{Qutrits}
\label{singlequtrits}

There are eight generators of SU($3$), usually represented in the form of 
Gell-Mann matrices. The Cartan sector is restricted to two diagonal matrices 
and the other six elements of the algebra are nondiagonal. Therefore, the 
SU($3$) transformations are determined by six parameters in the nondiagonal 
sector and two in the Cartan sector. In order to focus on the fractional 
phases and the role played by the state purity, we shall restrict our 
study to transformations restricted to the Cartan sector. The nondiagonal 
parameters only bring geometric complexity, without much additional insight 
into the fractional phase structure. 

First, we assume the qutrit basis is set to render diagonal the initial 
density matrix $\rho_{\,0}\,$. In terms of the two diagonal Gell-Mann matrices 
(apart from a slightly different normalization) we can parametrize the 
unit purity vector as $\mathbf{\hat{q}}=(\cos\theta,\sin\theta,0,\dots,0)\,$. In 
this parametrization, the density matrix for the initial state becomes 
\begin{eqnarray}
\rho_{\,0} &=& \frac{\mathbb{1}}{3} + q\,\sqrt{\frac{2}{3}}\,\left(
\cos\theta\,H_1 +\sin\theta\,H_2 \right)\;,
\label{rho0qutrit}\\
&=& \frac{\mathbb{1}}{3} + \frac{2\,q}{3}
\left[
\begin{matrix}
\cos\left(\theta + \frac{2\pi}{3}\right) & 0 & 0 \\
0 & \cos\left(\theta + \frac{4\pi}{3}\right) & 0 \\
0 & 0 & \cos\theta \\
\end{matrix}
\right]
\nonumber
\end{eqnarray}
where 
\begin{eqnarray}
H_1 &=& - \frac{1}{\sqrt{6}}\,\left[
\begin{matrix}
1 & 0 & 0 \\
0 & 1 & 0 \\
0 & 0 & -2 \\
\end{matrix} \right]\;,
\nonumber\\
H_2 &=& - \frac{1}{\sqrt{2}}\,\left[
\begin{matrix}
1 & 0 & 0 \\
0 & -1 & 0 \\
0 & 0 & 0 \\
\end{matrix} 
\right]\;. 
\label{gellmann}
\end{eqnarray}
Then, the diagonal parameters are 
\begin{eqnarray}
x_0 &=& \sqrt{\frac{2}{3}}\,\cos\left(\theta + \frac{2\pi}{3}\right)\;,
\nonumber\\
x_1 &=& \sqrt{\frac{2}{3}}\,\cos\left(\theta + \frac{4\pi}{3}\right)\;,
\\
x_2 &=& \sqrt{\frac{2}{3}}\,\cos\theta\;.
\nonumber
\end{eqnarray}
The density matrix eigenvalues belong to the interval $[0,1]$, what limits 
the possible values of $\theta\,$. For pure states ($q=1$), 
only the discrete values $0, 2\pi/3$ and $4\pi/3$ are allowed. For 
$q<1/2\,$, any value of $\theta$ gives a meaningful density 
matrix. Moreover, cyclic permutations of the basis vectors amount to 
transformations $\theta\rightarrow\theta+2\,n\,\pi/3$ ($n\in\mathbb{Z}$), 
and the noncyclic permutations can be achived by the same transformations 
followed by $\theta\rightarrow -\theta\,$. 
Therefore, without loss of generality, we can restrict our analysis to the domain 
$-\pi/3\leq\theta\leq\pi/3\,$. Other $\theta$ values simply amount to a permutation 
of the diagonal elements in $\rho_0\,$. Nevertheless, in order to ensure that all 
diagonal elements belong to the allowed interval $[0,1]$, we need to impose the 
restriction $\cos\left(\theta+2\,n\,\pi/3\right)\geq -1/2q\,$. 
Therefore, we arrive to $-\theta_0\leq\theta\leq\theta_0\,$, where  
\begin{equation}
\theta_0(q)= \left\{
\begin{matrix}
\cos^{-1}\left(-1/2q\right)-2\pi/3 & q\geq 1/2 \\
\pi/3 & q\leq 1/2 
\end{matrix}
\right.\;\;.
\label{theta0qutrit}
\end{equation}
For pure states ($q=1$) we are left with $\theta=0\,$.

We now assume a diagonal SU($3$) transformation 
\begin{equation}
\bar{U}(t)= \left[
\begin{matrix}
e^{i\,\chi_0(t)} & 0 & 0 \\
0 & e^{i\,\chi_1(t)} & 0 \\
0 & 0 & e^{i\,\chi_2(t)} \\
\end{matrix}
\right]\;, 
\label{ubarqutrit}
\end{equation}
where $\chi_2 = -(\chi_0 + \chi_1)\,$. 
In terms of the parameters characterizing $\bar{U}$ and $\rho_0\,$, 
the geometric phase becomes 
\begin{eqnarray}
\phi_g &=& \bar{\phi}_{\,tot} - 
\frac{2q}{3}\left[\chi_0\,\cos\left(\theta+\frac{2\pi}{3}\right)+ 
\chi_1\,\cos\left(\theta+\frac{4\pi}{3}\right)\right.
\nonumber\\
&+& \left. \chi_2\,\cos\theta\frac{}{}\,\right]\;,
\label{phigqutrit}
\end{eqnarray}
where
\begin{eqnarray}
\bar{\phi}_{\,tot} &=& 
\arg\left\{
e^{i\chi_0}\,\left[\frac{1}{3}+\frac{2q}{3}\,\cos\left(\theta+\frac{2\pi}{3}\right)\right]\right.
\nonumber\\
&+& \left. e^{i\chi_1}\,\left[\frac{1}{3}+\frac{2q}{3}\cos\left(\theta+\frac{4\pi}{3}\right)\right]\right. 
\nonumber\\
&+& \left. e^{i\chi_2}\,\left[\frac{1}{3}+\frac{2q}{3}\,\cos\theta\frac{}{}\right]
\right\}\;.
\label{phitotsinglequtrit}
\end{eqnarray}
We save this expression for our numerical investigation of the fractional phases 
acquired by entangled qutrits. 

%%%%%%%%%%%%%%%%%%%%%%%%%%%%%%%%%%%%%%%%%%%%%%%%%%%%%%%%%%%%%%%%%%%%%%%%%%%%%%%%%%%%%%%%%%%%%%%

\section{Fractional topological phases on entangled qudits}
\label{twoqudit}

We now turn to the main subject of this article, the fractional phases 
acquired by entangled qudits when subjected to local unitary operations.
We shall restrict our analysis to overall pure states.  
However, the results of the previous section will naturally extend to 
combined quantum systems, with the special role of entanglement, as 
measured by the purity of the partial density matrices. 

\subsection{Singular value decomposition}

We consider a two-qudit system with dimensions $d_A$ and $d_B\,$ ($d_A \leq d_B\,$). 
Let 
\begin{equation}
\ket{\psi}=\sum_{i=1}^{d_A}\sum_{j=1}^{d_B} \alpha_{ij}\ket{ij}
\label{psialpha}
\end{equation}
be the most general two-qudit {\it pure } state. We shall represent it by the 
$d_A\times d_B$ rectangular matrix $\mathbf{\alpha}$ whose elements are the coefficients 
$\alpha_{ij}$. With this notation, the associated norm becomes 
$\langle\psi|\psi\rangle=\mathrm{Tr}(\mathbf{\alpha^{\dagger}\alpha})=1$, and 
the scalar product between two states is 
$\langle\phi|\psi\rangle=\mathrm{Tr}(\mathbf{\beta^{\dagger}\alpha})$, where 
$\mathbf{\beta}$ is the $d_A\times d_B$ matrix representing $\ket{\phi}$ in the chosen basis. 
In order to characterize a general vector in the Hilbert space, 
we note that any invertible matrix admits a singular value decomposition
$\mathbf{\alpha}=e^{i\phi}\,S_A\,K\,S_B^T$, where $S_j \in$ SU($d_j$) ($j=A,B$), 
$K$ is a diagonal $d_A\times d_B$ rectangular matrix with real positive entries 
\begin{eqnarray}
K &=&\left[
\begin{matrix}
Q & \mathbf{0}_{AB}\end{matrix}
\right]\;,
\label{K}
\end{eqnarray}
and $Q$ is a $d_A\times d_A$ Hermitian diagonal square matrix 
($K_{\alpha\alpha}=Q_{\alpha\alpha}\in\mathbb{R}\,$, with $0\leq\alpha\leq d_A-1$). 
Here $\mathbf{0}_{AB}\,$ is a matrix of order $d_A\times(d_B-d_A)\,$, 
with all entries equal to zero. 
The normalization condition implies $\mathrm{Tr}[Q^2]=1\,$. 

Under local unitary operations $U_A(t)$ and $U_B(t)\,$, the coefficient 
matrix is transformed according to
$\mathbf{\alpha}(t)=U_A(t)\,\alpha(0)\,U_B^T(t)\;.$
It can be readily seen that this kind of transformation preserves 
the singular decomposition and can be represented separately in 
each sector of the coefficient matrix: 
\begin{equation}
\mathbf{\alpha}(t)=e^{i\phi (t)}\,S_A(t)\,K\,S_B^T(t)\;,
\label{alphat}
\end{equation}
where $S_j(t)=\bar{U}_j(t)\,S_j(0)\,$, $\bar{U}_j$ is the SU($d_j$) 
part of the corresponding local unitary operation: 
$U_j(t)=e^{i\phi_j(t)}\,\bar{U}_j(t)\,$, and 
$\phi (t)=\phi_0+\phi_A(t)+\phi_B(t)\,$. Therefore, we identify the 
transformation in three  sectors of the matrix structure: an explicit 
phase transformation $\phi_0\mapsto\phi(t)$, and two local evolutions 
$S_j(0)\mapsto S_j(t)$ ($j=A,B$) in SU($d_j$). The $K$-sector 
remains invariant under local unitary operations. 
At this point, we would like to note that the singular value decomposition is not 
unique, since different choices of $S_j$ may result 
in the same $\alpha\,$. However, this is not 
a problem, as long as one picks up any choice compatible with the 
initial coefficient matrix $\alpha(t=0)\,$. Then, the time evolution 
will be uniquely determined by the local unitary operations applied to the 
qudits.

In order to make a connection with the results of the previous section, it 
is important to unveil the physical meaning of the elements that participate 
in the singular value decomposition. It is easy to show that the reduced density 
matrices for qudits A and B are, respectively, 
\begin{equation}
\rho_A=\alpha\,\alpha^\dagger = S_A\,K\,K^{\dagger}\,S_{A}^{\dagger}
\label{defrhoA}
\end{equation}
and 
\begin{equation}
\rho_B=\left(\alpha^\dagger\,\alpha\right)^T = S_B\,K^{\dagger}\,K\,S_{B}^{\dagger}\;.
\label{defrhoB}
\end{equation}
Since $K$ is a $d_A\times d_B$ rectangular matrix with $d_A\leq d_B\,$, one imediately sees 
that 
\begin{eqnarray}
K\,K^{\dagger}=Q^2\;,
\end{eqnarray}
and
\begin{eqnarray}
K^{\dagger}\,K&=&\left[
\begin{matrix}
Q^2 & \mathbf{0}_{AB}\\
\mathbf{0}_{BA} & \mathbf{0}_{BB}
\end{matrix}
\right]\;,
\label{rhoB}
\end{eqnarray}
where $\mathbf{0}_{BA}$ and $\mathbf{0}_{BB}$ are matrices of order 
$(d_B-d_A)\times d_A$ and $(d_B-d_A)\times (d_B-d_A)\,$, respectively, with all 
entries equal to zero. When $d_A=d_B$ we obtain $K^{\dagger}\,K=K\,K^{\dagger}=Q^2\,$. 
It will be useful to parameterize our matrices in terms of the 
generators $\{T^j_{\alpha}\}$ of SU($d_j$), normalized as 
$\mathrm{Tr}\,[T^j_{\alpha}\,T^j_{\beta}]=\delta_{\alpha\beta}\,$. They 
can be separated into the Cartan subalgebra generators 
$H^j_{\alpha}$ with $[H^j_{\alpha},H^j_{\beta}]=0\,$, and the 
nondiagonal generators $P^j_{\alpha}\,$. 
Since $Q\,$ is a diagonal matrix, we can write
\begin{equation}
K\,K^{\dagger} = Q^2 = \frac{\mathbb{1}}{d_A} + 
q_A\,\sqrt{\frac{d_A-1}{d_A}}\,\mathbf{\hat{q}}_A\cdot\mathbf{H}^A\;,
\label{Q2}
\end{equation}
and
\begin{equation}
K^{\dagger}\,K=\frac{\mathbb{1}}{d_B}+q_B\,\sqrt{\frac{d_B-1}{d_B}}\,\mathbf{\hat{q}}_B\cdot\mathbf{H}^B\;.
\label{K2}
\end{equation}

Here $\mathbf{q}_j=q_j\,\mathbf{\hat{q}}_j\in \mathbb{R}^{d_j^2-1}$ is the purity vector 
associated with the reduced density matrix of qudit $j$ ($j=A,B$). 
As before, its absolute value is related to the state purity of qudit $j$ through 
$\mathrm{Tr}[\rho_j^2(0)]=q_j^2+(1-q_j^2)/d_j\,$. 
From $\mathrm{Tr}[(K\,K^{\dagger})^2]=\mathrm{Tr}[(K^{\dagger}\,K)^2]=
\mathrm{Tr}[Q^4]\,$, one easily shows that the norms of the two purity vectors are 
related by
\begin{equation}
q_B^2\,\frac{d_B-1}{d_B}=q_A^2\,\frac{d_A-1}{d_A}+\frac{d_B-d_A}{d_A\,d_B}\;.
\end{equation}
Moreover, the components of $\mathbf{\hat{q}}_B$ are not all independent 
because of the zeros on the diagonal of $K^\dagger\,K$ as given by Eq.(\ref{rhoB}). 
In fact, there will be only $d_A$ independent elements in $\mathbf{\hat{q}}_B$. 
Of course, if $d_A=d_B\,$, then $K\,K^{\dagger}=K^{\dagger}\,K=Q^2$ and the same 
generators as well as the same purity vector can be used for both qudits. 

It is now important to identify the following invariants under local unitary evolutions: 
$\mathrm{Tr}[\rho_j^{\, p}]$, $p=1,\dots, d$, where $\rho_j$ is the reduced density matrix 
with respect to qudit $j\,$.
In fact, the invariants are $j$-independent since one easily shows that 
$\mathrm{Tr}[\rho_A^{\, p}]=\mathrm{Tr}[\rho_B^{\, p}]=\mathrm{Tr}[Q^{2p}]\,$. 
The first one ($p=1$) is simply the 
norm of the state vector, as already stated. The second invariant is related to the 
{\it I-concurrence} of a two-qudit pure quantum state \cite{Iconc} 
\begin{eqnarray}
C=\sqrt{2(1-\mathrm{Tr}\,\rho_j^2)}=\sqrt{1-q_A^2\,}\,C_m\;, 
\end{eqnarray}
where
\begin{eqnarray}
C_m\equiv\sqrt{2\left(\frac{d_A-1}{d_A}\right)\,}\;, 
\end{eqnarray}
is the I-concurrence for maximally entangled states. The invariance 
of $C$ expresses the well 
known fact that entanglement is not affected by local unitary operations. 
The $p=d_A$ invariant can be rewritten in terms of the former and 
${\cal D}=|\det{Q}|$. In particular, for qubits we have 
$C=2\, {\cal D}$. 
In order to exploit the role played by these invariants in the geometric 
phase, we shall make them explicit in the expression of $Q^2$ by expressing the 
norm of the purity vector in terms of the I-concurrence, giving 
\begin{eqnarray}
Q^2 &=& \frac{\mathbb{1}}{d_A} + \sqrt{\frac{C_m^2-C^2}{2}}\,\,\mathbf{\hat{q}}_A\cdot \mathbf{H}^A
\\
&=& \frac{\mathbb{1}}{d_A}+\sqrt{\frac{C_m^2-C^2}{2}}\,
\mathrm{diag}\left[x_0\ldots x_{d-1}\right]\;,
\nonumber
\label{Q22}
\end{eqnarray}
with $x_n=\bra{n}\,\mathbf{\hat{q}}_A\cdot \mathbf{H}^A\,\ket{n}\,$, 
$\sum_n x_n=0$ and $\sum_n x_n^2=1\,$.

%%%%%%%%%%%%%%%%%%%%%%%%%%%%%%%%%%%%%%%%%%%%%%%%%%%%%%%%%%%%%%%%%%%%%%%%%%%%%%%%%%%%%%%%%%%%%%%

\subsection{Fractional phases}

Following \cite{smukunda,smukunda2}, we shall define as cyclic those evolutions for 
which the initial and final state vectors are related by a global 
phase factor: $\mathbf{\alpha^{\prime}}=e^{i\theta}\mathbf{\alpha}\,$, 
thus defining a closed path in the projective space of states ${\cal P}$. 
In other words, the final state of a cyclic evolution is physically 
equivalent to the initial one. The geometric phase acquired by a time 
evolving pure state $\mathbf{\alpha}(t)$ is given by 
\begin{eqnarray}
\phi_g &=& \arg{\langle\psi(0)|\psi(t)\rangle} + 
i\int dt \,\,\langle\psi(t)|\dot{\psi}(t)\rangle 
\label{phig}\\
&=& \arg\left\{\mathrm{Tr}\left[\mathbf{\alpha^{\dagger}}(0)\mathbf{\alpha}(t)\right]\right\} 
+ i\int dt\,\, \mathrm{Tr}\left[\mathbf{\alpha^{\dagger}}(t)\dot{\mathbf{\alpha}}(t)\right]
\nonumber\;,
\end{eqnarray}
which corresponds to the total phase 
\begin{equation}
\phi_{\,tot}\equiv\arg\left\{\mathrm{Tr}
\left[\mathbf{\alpha^{\dagger}}(0)\,\mathbf{\alpha}(t)\right]\right\}\;, 
\end{equation}
minus the dynamical phase. We now use 
the singular value decomposition to investigate the contribution originated 
from each sector of the coefficient matrix. First, we can write the 
total phase as 
\begin{equation}
\phi_{\,tot}=\phi(t)-\phi(0)+\bar{\phi}_{tot}\;,
\label{phitot1} 
\end{equation}
where
\begin{eqnarray}
\bar{\phi}_{\,tot}&\equiv&\arg\left\{\mathrm{Tr}\left[
\mathbf{\alpha^{\dagger}}(0)\,\bar{U}_A(t)\,\mathbf{\alpha}(0)\bar{U}_B^T(t)\right]\right\}\;,
\label{phibartot}
\end{eqnarray}
is the contribution brought by the SU($d_j$) sectors. 
We can investigate the dynamical phase using the 
singular value decomposition; using Eq.(\ref{alphat}) we obtain
\begin{eqnarray}
\mathrm{Tr}\left[\mathbf{\alpha}^{\dagger}\dot{\mathbf{\alpha}}\right]&=&
i\,\dot{\phi} + 
\mathrm{Tr}\left[S_A^{\dagger}\dot{S}_A\,K\,K^{\dagger} + 
K^{\dagger}\,K\,S_B^{\dagger}\dot{S}_B\right]
\nonumber\\
&=& i\,\dot{\phi} + 
\mathrm{Tr}\left[\rho_A(0)\bar{U}_A^{\dagger}\dot{\bar{U}}_A + 
\rho_B(0)\bar{U}_B^{\dagger}\dot{\bar{U}}_B\right]
\,.
\nonumber\\
\label{tralphadot0}
\end{eqnarray}
Note that the trivial phase evolution $\phi(t)$ cancels out 
when Eqs.(\ref{phitot1}) and (\ref{tralphadot0}) are used 
in the geometric phase expression, so that we are left with 
\begin{eqnarray}
\phi_g=\bar{\phi}_{tot}-\int \mathrm{Tr}\left[\rho_A(0)\bar{U}_A^{\dagger}\dot{\bar{U}}_A + 
\rho_B(0)\bar{U}_B^{\dagger}\dot{\bar{U}}_B\right]\,dt
\,.
\label{phigrho}
\end{eqnarray}

The reduced density matrices at $t=0$ can also be expanded in terms of 
the identity matrix and the generators of SU($d_j$) as 
\begin{equation}
\rho_j(0)=\frac{\mathbb{1}}{d_j} + 
q_j\,\sqrt{\frac{d_j-1}{d_j}}\,\mathbf{\hat{q}}_j^{\prime}\cdot\mathbf{T}^j\;.
\label{rhoj0}
\end{equation}
Note that $\mathbf{\hat{q}}_j$ and $\mathbf{\hat{q}}_j^{\prime}$ are 
connected by an initial rotation in 
$\mathbb{R}^{d_j^2-1}\,$, which is contained in the adjoint representation of $HU(d)$. This is 
determined by $S_j(0)\,$ through 
$S_j(0)\left(\mathbf{\hat{q}}_j\cdot\mathbf{H}^j\right)S_j^{\dagger}(0)=
\mathbf{\hat{q}}_j^{\prime}\cdot\mathbf{T}^j\,$. In particular, if the 
local bases are chosen in order to diagonalize the initial 
two-qudit density matrix, then we can make $S_j(0)=\mathbb{1}$ 
and $\mathbf{\hat{q}}_j^{\prime} = \mathbf{\hat{q}}_j\,$.  

Let us consider a cyclic evolution over the time interval $T\,$, 
$\mathbf{\alpha}(T)=e^{i\theta}\mathbf{\alpha}(0)\,$. 
By defining the local velocity vectors $\mathbf{u}_j$ according to 
$\bar{U}_j^{\dagger}\dot{\bar{U}}_j=i\,\mathbf{u}_j\cdot \mathbf{T}^{j}$ 
and using the orthogonality condition for the generators, we arrive at
\begin{eqnarray}
\phi_g &=& \bar{\phi}_{tot} - \sqrt{\frac{C_m^2 - C^2}{2}}\,
\oint \mathbf{\hat{q}}_A^{\prime}\cdot\mathbf{dx}_A
\label{phig2} \nonumber\\
&-& 
\sqrt{\frac{C_m^2-C^2}{2}+\frac{d_B-d_A}{d_A\,d_B}\,}\,
\oint \mathbf{\hat{q}}_B^{\prime}\cdot\mathbf{dx}_B
\;, 
\end{eqnarray}
where $\mathbf{dx}_j=\mathbf{u}_j\,dt\,$. 

Now, let us analyze the total phase. Since $\mathbf{\alpha}(T)=e^{i\phi(T)}\,S_A(T)\,K\,S_B^T(T)\,$, and 
the $K$ sector is time independent, the global phase $\theta$ acquired by the 
two-qudit system is composed by a trivial phase evolution plus the contributions 
from the SU($d_j$) sectors, 
\begin{equation}
\theta=\phi(T)-\phi(0)+\theta_A+\theta_B\;, 
\end{equation}
where $S_j(T)=e^{i\theta_j}S_j(0)\,$ ($j=A,B$). However, 
$\det{S_j(T)}=e^{id_j\theta_j}\det{S_j(0)}$, 
and since these evolutions are closed in the space of SU($d_j$) matrices, we arrive at
\begin{equation}
\theta_j=2\pi\,\frac{n_j}{d_j}\;,
\label{thetafrac}
\end{equation}
where $n_j\in\mathbb{Z}\,$. Therefore, the trivial phase is canceled 
by the integral term and only fractional phase values can arise 
from the SU($d_j$) sectors 
\begin{equation}
\bar{\phi}_{\,tot}=2\pi\,\left(\frac{n_A}{d_A}+\frac{n_B}{d_B}\right)\;.
\label{theta}
\end{equation}
Then, the geometric phase acquired in the cyclic evolution becomes
\begin{eqnarray}
\phi_g &=& 2\,\pi\,\left(\frac{n_A}{d_A}+\frac{n_B}{d_B}\right) - 
\sqrt{\frac{C_m^2 - C^2}{2}}\,\oint \mathbf{\hat{q}}_A^{\prime}\cdot\mathbf{dx}_A
\label{phigfrac} 
\nonumber\\
&-& 
\sqrt{\frac{C_m^2-C^2}{2}+\frac{d_B-d_A}{d_A\,d_B}\,}\,
\oint \mathbf{\hat{q}}_B^{\prime}\cdot\mathbf{dx}_B
\;. 
\end{eqnarray}
Eq.(\ref{phigfrac}) evidences the roles played by entanglement and the 
dimensions of the qudit Hilbert spaces. When $d_A=d_B\,$, it reduces to 
the result reported in ref.\cite{fracuff} with $n=n_A+n_B\,$. In this case, 
as anticipated in section \ref{geosingle}, for maximally entangled states 
the partial traces give a completely mixed density matrix for each qudit, 
so that only fractional geometric phases are allowed. Here, since the 
complete two-qudit state considered is pure, these fractional phases can 
be evidenced through conditional interference \cite{fracuffufmg} when 
the qudits are locally operated with SU($d$) transformations. 

%%%%%%%%%%%%%%%%%%%%%%%%%%%%%%%%%%%%%%%%%%%%%%%%%%%%%%%%%%%%%%%%%%%%%%%%%%%%%%%%%%%%%%%%%%%%%%%

\subsection{The two-qudit Cartan sector}
\label{Cartan2}

Similarly to section \ref{Cartan}, the local unitary evolutions can be 
decomposed into the Cartan ${\rm U}(1)^{d_j-1}$ sector and
the coset manifold ${\rm SU}(d_j)/{\rm U}(1)^{d_j-1}$. This decomposition 
gives rise to two separate integral terms, as in 
Eq.(\ref{phigsinglequdit}). Let us assume that the local basis is chosen so as to make 
the matrix $\mathbf{\alpha}(0)\,$ diagonal. In this representation, $S_j(0)=\mathbb{1}$ and 
the reduced density matrices at $t=0$ are simply 
\begin{eqnarray}
\rho_A(0)&=&\frac{\mathbb{1}}{d_A}+\sqrt{\frac{C_m^2-C^2}{2}\,}\,\mathbf{\hat{q}}_A\cdot\mathbf{H}^A\;, 
\label{rhoABCartan}\\
\rho_B(0)&=&\frac{\mathbb{1}}{d_B}+\sqrt{\frac{C_m^2-C^2}{2}+
\frac{d_B-d_A}{d_A\,d_B}\,}\,\mathbf{\hat{q}}_B\cdot\mathbf{H}^B\;. 
\nonumber
\end{eqnarray}

Now, we can employ the decomposition in Eq.(\ref{UVbarqudit}) 
\begin{eqnarray}
\bar{U}_j &=& 
\bar{V}_j\,\exp\left(i\,\mathbf{h}_j\cdot \mathbf{H}^j\right)\;,
\label{UVbarquditj}
\end{eqnarray}
($j=A,B$) and separate the Cartan sectors for each qudit evolution.  
The velocity vectors $\mathbf{u}_A$ and $\mathbf{u_B}$ can be decomposed as in Eq.(\ref{decomp-u}), 
$\mathbf{u}_j=\mathbf{v}^{\,\prime}_{j\,\bot} + \mathbf{v}_{j\,\|} + \dot{\mathbf{h}}_{j}\,$. 
Since the reduced density matrices are written in a diagonal representation, 
$\hat{\mathbf{q}}_{j}\cdot\mathbf{v}^{\,\prime}_{j\,\bot}=0$ so that only $\mathbf{v}_{j\,\|}$ and 
$\dot{\mathbf{h}}_{j}$ will contribute to the integral term in the geometric phase. The contribution 
from $\mathbf{\dot{h}}_{j}$ is path independent (holonomic).
The path dependent (nonholonomic) contribution from $\mathbf{v}_{j\,\|}$ captures the geometric 
nature of the evolution in $\mathrm{SU}(d_j)/\mathrm{U}(1)^{d_j-1}$. 

Suppose that, at time $\bar{t}$, a partially cyclic evolution 
occurs. Then, we have 
\begin{eqnarray}
\phi_g &=& \bar{\phi}_{\,tot} - \sqrt{\frac{C_m^2 - C^2}{2}}\,
\left[\mathbf{\hat{q}}_A\cdot\mathbf{h}_A(\bar{t})+\Phi_A\right]
\label{phig2partial}\\
&-& 
\sqrt{\frac{C_m^2-C^2}{2}+\frac{d_B-d_A}{d_A\,d_B}\,}\,
\left[\mathbf{\hat{q}}_B\cdot\mathbf{h}_B(\bar{t})+\Phi_B\right]\;,
\nonumber 
\end{eqnarray}
where
\begin{eqnarray} 
\Phi_j &=& \oint \,\, \mathbf{\hat{q}}_j\cdot\mathbf{dx}_{j\|}\;, 
\label{Omegapj} 
\end{eqnarray} 
and $\mathbf{dx}_{j\|}=\mathbf{v}_{j\|}\,dt$ ($j=A,B$). 

For partially cyclic evolutions, the same argument leading to Eq.(\ref{Ub}), here implies
\begin{eqnarray}
\bar{U}_j(\bar{t}) &=& 
\exp\left(i\,\mathbf{h}_j(\bar{t})\cdot \mathbf{H}^j\right)\;,
\label{Ub2}
\end{eqnarray}
\begin{eqnarray}
\bar{\phi}_{\,tot}&\equiv&\arg\left\{\mathrm{Tr}\left[
\mathbf{\alpha^{\dagger}}(0)\,e^{i\,\mathbf{h}_A(\bar{t})\cdot \mathbf{H}^A}\,\mathbf{\alpha}(0)
e^{i\,\mathbf{h}_B(\bar{t})\cdot \mathbf{H}^B}\right]\right\}\;.\nonumber \\
\label{fbt} 
\end{eqnarray}
If in addition the evolution is cyclic, then the condition, 
\begin{eqnarray}
\bar{U}_j(T) &=& 
\exp\left(i\,\mathbf{h}_j(T)\cdot \mathbf{H}^j\right) = \exp i (2\pi\,n_j/d_j)\,\mathbb{1}\;,
\nonumber \\
\label{Cond}
\end{eqnarray}
must be satisfied, and the geometric phase is given by Eq.(\ref{phig2partial}), 
with $\bar{t} \to T$, and $\bar{\phi}_{\,tot}$ given by the fractional values in 
Eq.(\ref{theta}); for qudits with equal dimensions $d_A=d_B=d\,$, 
\begin{eqnarray}
\phi_g &=& \frac{2\pi}{d}\,\left(n_A + n_B\right) - \sqrt{\frac{C_m^2 - C^2}{2}}\,
\mathbf{\hat{q}}\cdot\left[\mathbf{h}_A(T)+\mathbf{h}_B(T)\right]
\nonumber\\
&-& 
\sqrt{\frac{C_m^2-C^2}{2}\,}\,\left(\Phi_A+\Phi_B\right)\;.
\label{phig2partialequald}
\end{eqnarray}

For given $n_j$ values, there is a discrete set $\{ \mathbf{h}_j(T)\}_{n_j}$ 
of solutions to Eq.(\ref{Cond}),
forming a lattice in $\mathbb{R}^{d_j-1}\,$, which must be attained by $\mathbf{h}_j(t)$ 
in order to produce closed paths in the projective space of states ${\cal P}$. 
Those cyclic evolutions $\{ \mathbf{h}_j(t)\}_0$ characterized by $n_j=0$ also describe 
closed paths in $SU(d_j)$, 
so they are topologically trivial, as $SU(d_j)$ is simply connected. 
On the other hand, take for example cyclic evolutions $\{ \mathbf{h}_j(t)\}_1$, 
characterized by $n_j=1$.  
They correspond to  topologically {\it nontrivial} closed paths in ${\cal P}$ as, i) 
they are open in $SU(d_j)$, so the triviality of closed paths in $SU(d_j)$ 
does not apply in this case ii) the lattices $\{ \mathbf{h}_j(T)\}_0$ and 
$\{ \mathbf{h}_j(T)\}_1$ are different and, iii) the general condition (\ref{Cond}), 
to keep the paths closed in ${\cal P}$, lead to discrete possibilities, with no 
solutions continuously interpolating the $n_j=0$ and $n_j=1$ lattices.

Note that closed paths with a fixed base point, and open paths with fixed endpoints, 
are fundamental elements to characterize the topological structure of a manifold. 
As is well-known, the consideration of equivalence classes of closed paths, and 
the natural product based on their composition, 
leads to the first homotopy group.

For example, consider an evolution that interchanges a pair of anyons. This would 
correspond to a closed path in the configuration space of indistinguishable particles 
on the plane, as well as a closed path in the projective space of two-anyon states. 
To generate the fractional statistics phases, this type of evolution should be 
controlled. A similar physical content is contained in the necessary condition 
(\ref{Cond}) to generate closed paths in the projective space ${\cal P}$ for a 
qudit pair.

\subsection{Diagonal evolutions}

Consider two qudits with the same 
Hilbert space dimension $d$ that are locally operated by diagonal SU($d$) matrices 
$\bar{U}_j=\mathrm{diag}\left[e^{i\,\chi_{j0}}\ldots e^{i\,\chi_{j(d-1)}}\right]$ 
($j=A,B$) starting from the initial state 
\begin{eqnarray}
\ket{\psi(0)}=\sum_{n=0}^{d-1}\left(\frac{1}{d} + 
q\sqrt{\frac{d-1}{d}}\,x_n\right)^{1/2}\,\ket{n\,n}\;.
\end{eqnarray}  
In this case $\rho_A(0)=\rho_B(0)=Q^2\,$. The geometric phase reduces to 
\begin{eqnarray}
\phi_g = \bar{\phi}_{tot} - \sqrt{\frac{C_m^2-C^2}{2}}\,
\sum_{n=0}^{d-1} x_n\,\chi_{T\,n}\;,
\label{phig2quditdiagonal}
\end{eqnarray} 
where $\chi_{T\,n}=\chi_{A\,n}+\chi_{B\,n}\,$. 
Since the coefficient matrix and the local operations are diagonal,
the evolution is partially cyclic at any time $t$. Then, we can use 
Eq.(\ref{fbt}) to obtain
\begin{eqnarray}
\bar{\phi}_{tot}=\arg\left\{\sum_{n=0}^{d-1}\,
\left(\frac{1}{d} + \sqrt{\frac{C_m^2-C^2}{2}}\,x_{\,n}\right)
\,e^{i\,\chi_{T\,n}}\right\}\;.
\nonumber\\ 
\label{phitot2quditdiagonal}
\end{eqnarray}
We note that Eqs.(\ref{phig2quditdiagonal}) and (\ref{phitot2quditdiagonal}) 
are very similar to (\ref{phigsinglediagonal}) and (\ref{phitotsinglediagonal}). 
However, for diagonal transformations of entangled states (with $d_A=d_B$), 
the overall cyclic transformation can be composed by local noncyclic operations, 
since the total and geometric phase only depend on $\mathbf{h}_A(t)+\mathbf{h}_B(t)$. 
This fact is crucial for experimental investigations of the fractional phases and the 
role played by entanglement.

%%%%%%%%%%%%%%%%%%%%%%%%%%%%%%%%%%%%%%%%%%%%%%%%%%%%%%%%%%%%%%%%%%%%%%%%%%%%%%%%%%%%%%%%%%%%%%%

\section{Examples}
\label{examples}

\subsection{Qubits Revisited}
\label{2qubits}

As an illustration of the methods used in the previous section, we now consider a two-qubit 
system ($d_A=d_B=2$) initially prepared in the state
\begin{eqnarray}\ket{\psi(0)}=\frac{\sqrt{1+q\,}\,\ket{00}\,+\,\sqrt{1-q\,}\,\ket{11}}{\sqrt{2}}\;,
\label{psi2qubit}
\end{eqnarray}
with $0\leq q\leq 1\,$. Note that any two-qubit pure state can be cast in this 
form by a suitable local basis choice. In this case a single purity vector $\mathbf{q}$ 
can be used for both qubits. For the state given by Eq.(\ref{psi2qubit}), the concurrence is 
$C=\sqrt{1-q^2\,}\,$, and 
\begin{eqnarray} 
Q^2 &=& \left[
\begin{matrix}
\frac{1+q}{2} & 0\\
\\
0 & \frac{1-q}{2}
\end{matrix}
\right]
%\nonumber\\
= \frac{\mathbb{1}}{2}+\frac{\sqrt{1-C^{\,2}}}{2}\,\sigma_z\;.
\label{alpha2qubit}
\end{eqnarray}
Also, we may choose $S_A(0)=S_B(0)=\mathbb{1}\,$. 
The associated purity vector simply is $\mathbf{q}=(\sqrt{1-C^{\,2}\,},0,0)\,$ 
($\mathbf{T}= (1/\sqrt{2})\, (\sigma_z,\sigma_x,\sigma_y)$ for SU($2$)). 
Let us assume that these qubits evolve 
under local unitary operators $U_j(t)=e^{i\phi_j(t)}\,\bar{U}_j(t)\,$, 
($j=A,B$) where $\bar{U}_j$ is a SU($2$) matrix acting on qubit $j$ and $\phi_j$ is the 
corresponding global phase introduced by $U_j\,$. As in section \ref{geosinglequbits}, 
we can make 
\begin{eqnarray}
\bar{U}_j(\theta_j,\varphi_j,\chi_j) &=& 
\bar{V}_j(\theta_j,\varphi_j)\,e^{i\chi_j\sigma_z}\;,
\label{Ubarj}
\end{eqnarray}
and 
\begin{eqnarray}
\bar{V}_j(\theta_j,\varphi_j) &=& 
\exp\left({i\,\theta_j\,\,\mathbf{\hat{p}}_j\cdot\mathbf{P}_j}\right)
\nonumber\\
&=&\left[
\begin{matrix}
\cos\frac{\theta_j}{2} & i\,\sin\frac{\theta_j}{2}\,e^{-i\varphi_j}\\
i\,\sin\frac{\theta_j}{2}\,e^{i\varphi_j} & \cos\frac{\theta_j}{2}
\end{matrix}
\right]\;,
\label{Vbarj}
\end{eqnarray}
where $\mathbf{\hat{p}}_j=(0,\cos\varphi_j,\sin\varphi_j)\,$. 
Here $\varphi_j(t)$, $\theta_j(t)$, and $\chi_j(t)$ are time dependent 
real parameters with initial conditions $\varphi_j(0)=\theta_j(0)=\chi_j(0)=0\,$. 
As before, $\varphi_j(t)$ and $\theta_j(t)$ can be identified with angular coordinates 
on two separate Bloch spheres, one for each qubit. 
Thus, the velocity vector 
corresponding to each evolution is given by
${\rm\bf v}_j = (v_{h}^j, v_{p1}^j ,v_{p\,2}^j)$, where
\begin{eqnarray}
v_{h}^j &=&  \sqrt{2}\,\dot{\varphi}_j\,\sin^2\left(\frac{\theta_j}{2}\right)\nonumber \\
v_{p1}^j &=& \frac{1}{\sqrt{2}} \left(\dot{\theta}_j\,\cos\varphi_j-\dot{\varphi}_j\,\sin\theta_j\,\sin\varphi_j\right) 
\nonumber\\
v_{p\,2}^j &=& \frac{1}{\sqrt{2}}\left(\dot{\theta}_j\,\sin\varphi_j + 
\dot{\varphi}_j\,\sin\theta_j\,\cos\varphi_j\right)\;,
\end{eqnarray}
and the component of $\mathbf{u}$ along the Cartan direction is,
\begin{eqnarray}
v_{h}^j + \sqrt{2}\,\dot{\chi}_j\;.
\end{eqnarray}
Now, from eqs.(\ref{intusingle}) and (\ref{phigqubit2}), the geometric 
phase for a pair of qubits following a cyclic evolution under local 
unitary operations reduces to
\begin{eqnarray}
\phi_g &=& n\pi - \frac{\sqrt{1-C^{\,2}}}{2}\left(\Omega_A+\Omega_B\right)\;,
\label{phigqubit}
\end{eqnarray}
where $n=n_A+n_B\,$. This is a quite intuitive result in which we identify the topological 
contribution first predicted in ref.\cite{remy2}, and the sum of the usual solid 
angle contributions from both qubits weighted by entanglement. For maximally entangled 
states, only the two fractional values are left. 

It will be particularly interesting to investigate the geometric phase acquired under  
\textit{partially cyclic} evolutions. For these evolutions 
$\left(\theta_j(t),\varphi_j(t)\right)$ follows a closed path on the Bloch sphere, but 
$\chi_j(t)$ does not necessarily make a full cycle. 
In this case, the geometric phase becomes
\begin{eqnarray}
\phi_g &=& \arctan\left[\sqrt{1-C^{\,2}}\,\tan\left(\chi_A+\chi_B\right)\right] 
\nonumber\\
&-& \sqrt{1-C^{\,2}}\,\left(\chi_A + \chi_B + \frac{\Omega_A + \Omega_B}{2}\right)\;.
\label{phigpartial} 
\end{eqnarray}
For product states ($C=0$), the $\chi_j$ terms cancel out and give no 
contribution to the geometric phase, which is then determined by the individual solid angles 
enclosed in the separate Bloch spheres. As the concurrence increases, the solid angle 
contributions diminish and a net effect of the $\chi_j$ terms appears as a stepwise variation 
of the geometric phase as a function of $\chi_{T}=\chi_A + \chi_B\,$. For maximally 
entangled states ($C=1$), the solid angle contributions completely vanish and the 
stepwise evolution degenerates to a discontinuous jump from $0$ to $\pi\,$, the allowed 
fractional phases for qubits. This simple example illustrates the role played by entanglement 
in the way the geometric phase is built during the evolution. 

The result given by Eq.(\ref{phigpartial}) is a generalization of Eq.(9) in Ref.\cite{fracuff} 
for the case where both qubits are operated. It is also very similar to Eq.(\ref{phigqubit3}) 
in section \ref{geosinglequbits}, specially if we notice that $\sqrt{1-C^{\,2}\,}=q\,$. 
Of course, this similarity is not surprising once we realize that the partial density matrices of 
the entangled qubits ($\rho_A(0)=\rho_B(0)=Q^2\,$, see Eq.(\ref{alpha2qubit})) are identical to 
the single qubit mixed state considered in Eq.(\ref{rho0qubit}). Therefore, the entanglement 
signature on the geometric phase evolution is directly related to the purity of the partial 
traces of the two-qubit density matrix. However, there are two important differences between 
the two cases. First, since the two-qubit entangled state considered here is pure, we can 
expect the fractional phases to be experimentally observable. 
Second, the geometric phase acquired by the entangled qubits depends on 
$\chi_{T}=\chi_A + \chi_B\,$, which means that the overall cyclic transformation 
can be split into local noncyclic operations applied to the entangled qubits separately.

%%%%%%%%%%%%%%%%%%%%%%%%%%%%%%%%%%%%%%%%%%%%%%%%%%%%%%%%%%%%%%%%%%%%%%%%%%%%%%%%%%%%%%%%%%%%%%%

\subsection{Qutrits}

For the two-qutrit case we will restrict our analysis to local evolutions in the 
3$\times$3 Cartan sector of each qutrit $\bar{U}_A(t)\otimes\bar{U}_B(t)\,$, 
where 
\begin{equation}
\bar{U}_j(t)= \left[
\begin{matrix}
e^{i\,\chi_{j0}} & 0 & 0 \\
0 & e^{i\,\chi_{j1}} & 0 \\
0 & 0 & e^{i\,\chi_{j2}} \\
\end{matrix}
\right]\;,
\label{ubar2qutrit}
\end{equation}
with $j=A,B$ and $\chi_{j\,0}+\chi_{j\,1}+\chi_{j\,2}=0\,$.

\begin{figure}[h!]
\includegraphics[scale=0.31]{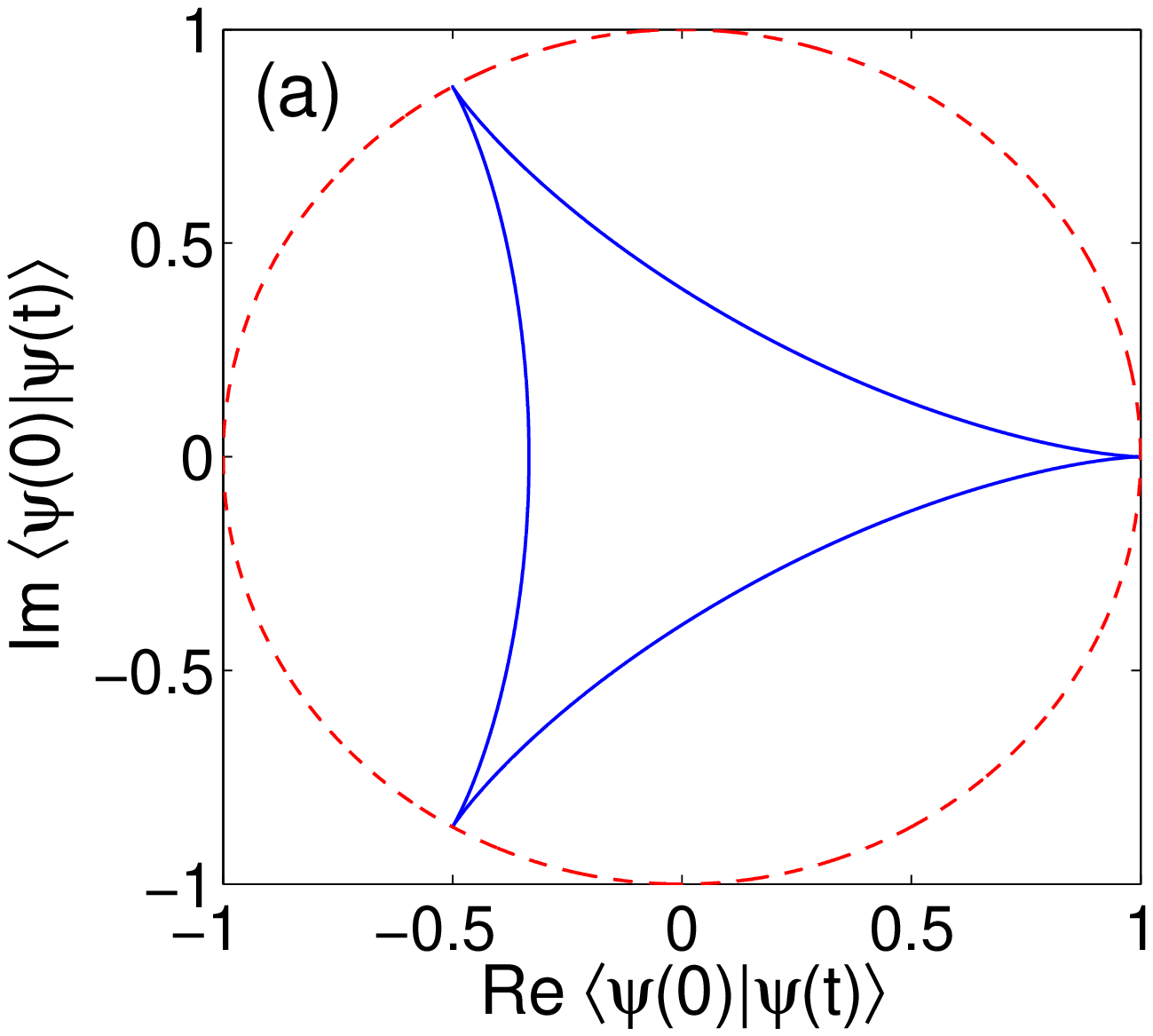}
\includegraphics[scale=0.31]{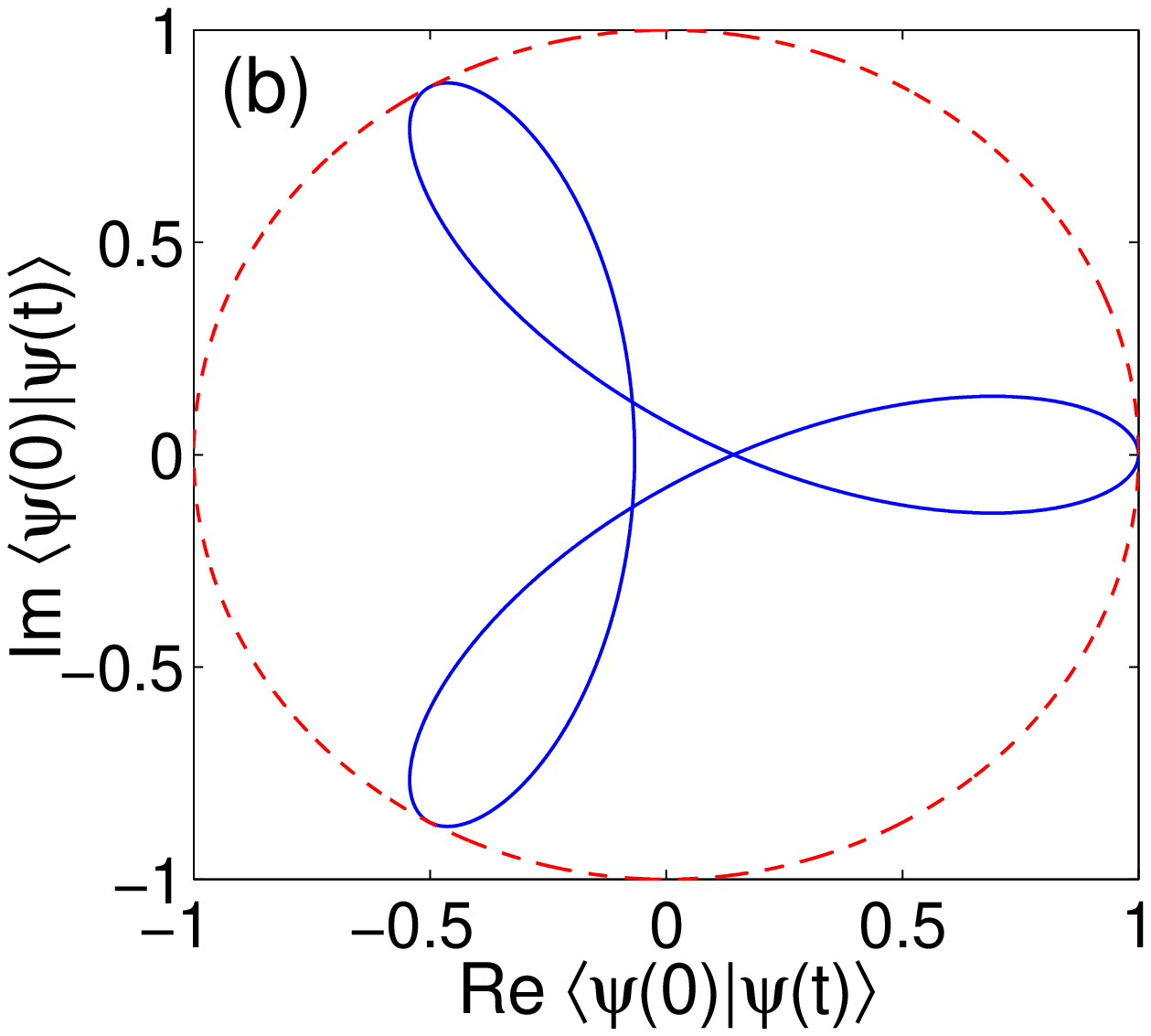}
\includegraphics[scale=0.31]{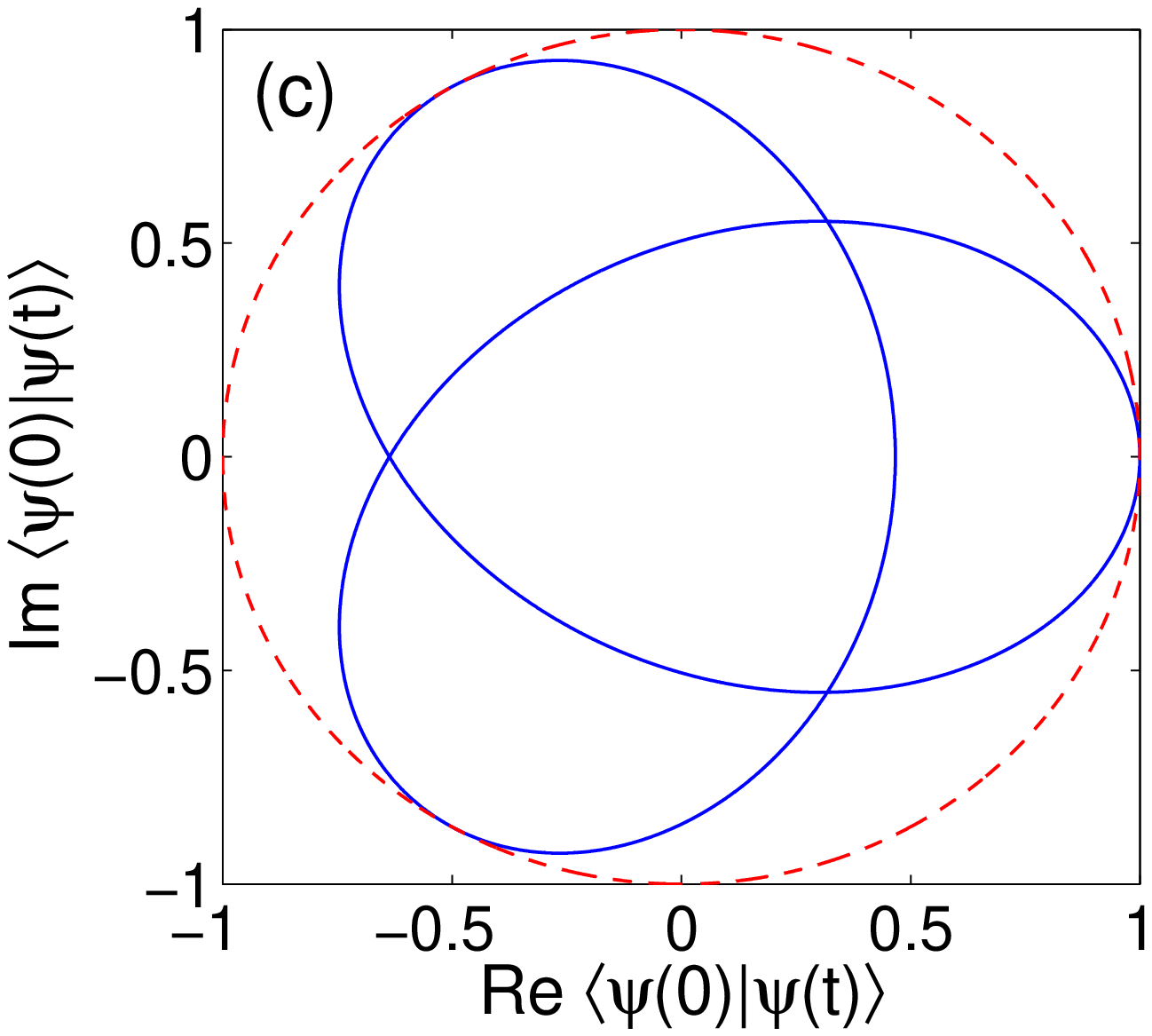}
\includegraphics[scale=0.31]{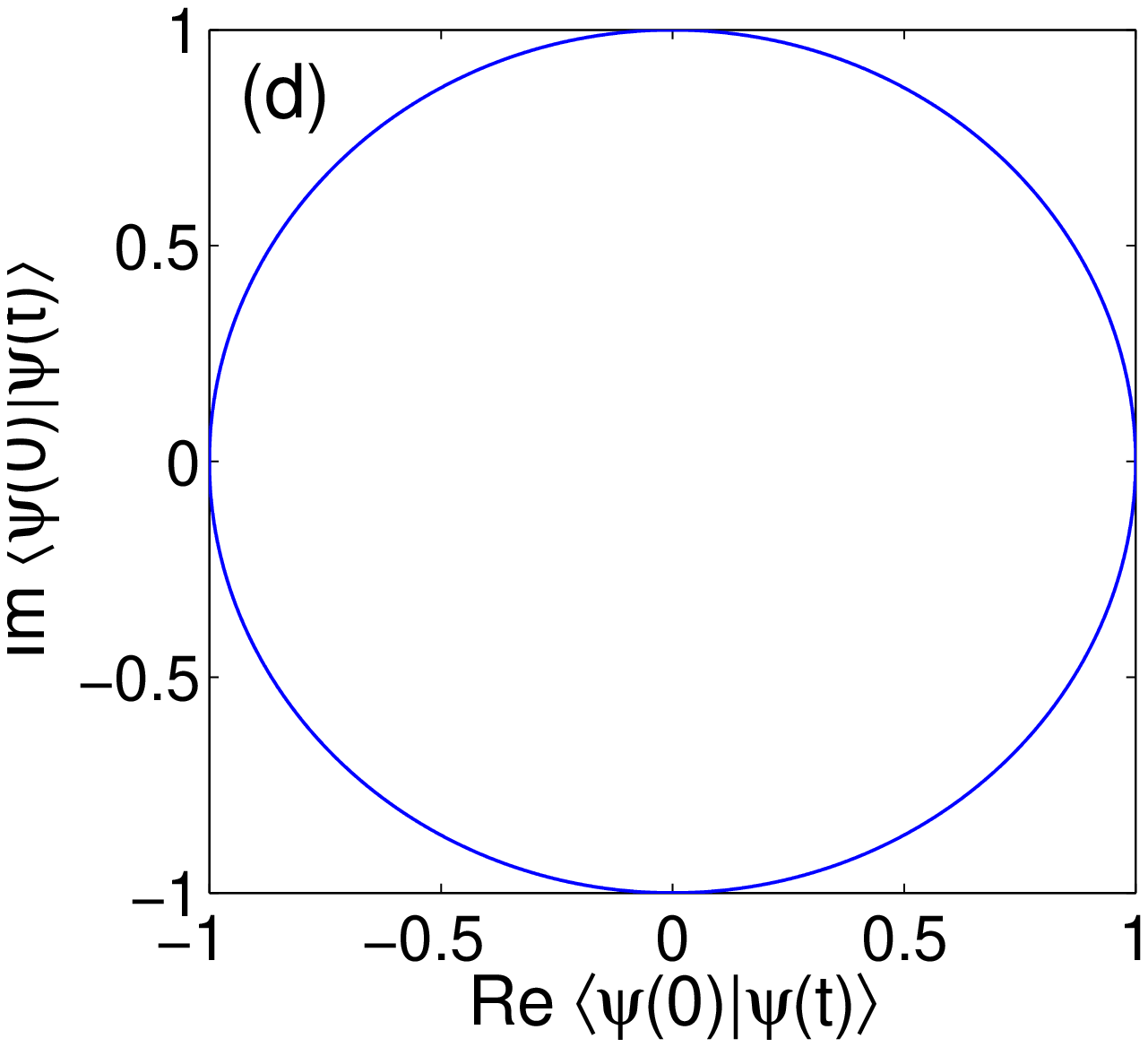}
\caption{\label{fig:sim1} Parametric plot of the quantum state overlap for a two-qutrit 
evolution given by Eqs.(\ref{evol2qutrit1}). 
(a) $q=0$ ($C=\sqrt{4/3}$), (b) $q=0.2\,$, (c) $q=0.6\,$, (d) $q=1$ ($C=0$). 
The unit circle is depicted in dashed red (online) for reference.}
\end{figure}

Let us suppose that the local basis is chosen so as to leave the initial 
two-qutrit pure state in the form
\begin{eqnarray}
\ket{\psi(0)}&=&\frac{1}{\sqrt{3}}\,
\left[\sqrt{1+2q\cos(\theta + 2\pi/3)}\,\ket{00}\right. 
\nonumber\\
&+&\sqrt{1+2q\cos(\theta + 4\pi/3)}\,\ket{11} 
\nonumber\\
&+&\left.\sqrt{1+2q\cos\theta)}\,\ket{22}\right]\;,
\label{2qutritstate}
\end{eqnarray}
where $0\leq q\leq 1\,$, $-\theta_0(q)\leq\theta\leq\theta_0(q)$ and 
$\theta_0(q)$ is given by (\ref{theta0qutrit}). The maximal concurrence 
for qutrits is $C_m=\sqrt{4/3}$ and the concurrence of state 
(\ref{2qutritstate}) is $C=C_m\,\sqrt{1-q^2}\,$. Since the coefficient 
matrix $\mathbf{\alpha}(0)$ and the local unitary operations are all 
diagonal, the calculation of the nontrivial total phase 
$\bar{\phi}_{tot}$ is significantly simplified as 
\begin{eqnarray}
\bar{\phi}_{tot}=\arg\left\{\mathrm{Tr}\left[Q^2\,\bar{U}_A(t)\,\bar{U}_B(t)\right]\right\}\;,
\end{eqnarray}
where
\begin{equation}
Q^2= \frac{\mathbb{1}}{3} + \frac{2\,q}{3}\,\left[
\begin{matrix}
\cos(\theta + \frac{2\pi}{3}) & 0 & 0 \\
0 & \cos(\theta + \frac{4\pi}{3}) & 0 \\
0 & 0 & \cos\theta \\
\end{matrix}
\right]\;.
\label{Q2qutrit}
\end{equation}
Also, the partial density matrices for qutrits A and B at $t=0$ are 
equal to $Q^2\,$. Therefore, the geometric phase 
becomes
\begin{eqnarray}
\phi_g &=& \bar{\phi}_{\,tot} - 
\frac{2q}{3}\left[\chi_{T0}\,\cos\left(\theta+\frac{2\pi}{3}\right)\right. 
\nonumber\\
&+& \left. \chi_{T1}\,\cos\left(\theta+\frac{4\pi}{3}\right)
+ \chi_{T2}\,\cos\theta\frac{}{}\,\right]\;,
\label{phig2qutrit}
\end{eqnarray}
where $\chi_{Tn}=\chi_{An}+\chi_{Bn}\,$, and 
the nontrivial total phase is 
\begin{eqnarray}
\bar{\phi}_{\,tot} &=& 
\arg\left\{
e^{i\chi_{T0}}\,\left[\frac{1}{3}+\frac{2q}{3}\,\cos\left(\theta+\frac{2\pi}{3}\right)\right]\right.
\nonumber\\
&+& \left. e^{i\chi_{T1}}\,\left[\frac{1}{3}+\frac{2q}{3}\cos\left(\theta+\frac{4\pi}{3}\right)\right]\right. 
\nonumber\\
&+& \left. e^{i\chi_{T2}}\,\left[\frac{1}{3}+\frac{2q}{3}\,\cos\theta\frac{}{}\right]
\right\}\;.
\label{phitot2qutrits}
\end{eqnarray}
Eqs.(\ref{phig2qutrit}) and (\ref{phitot2qutrits}) are very similar to 
the single qutrit result given by Eqs.(\ref{phigqutrit}) and 
(\ref{phitotsinglequtrit}). However, the diagonal phase shifts $\chi_n$ 
are replaced by the total phase shifts $\chi_{Tn}\,$, showing the nonlocal 
character of the geometric phase. For maximally entangled states 
($q=0\Rightarrow C=C_m$), the integral term vanishes, so that 
$\phi_g=\bar{\phi}_{tot}\,$. In Fig.(\ref{fig:sim1}) we show a 
parametric plot of the overlap $\braket{\psi(0)}{\psi(t)}$ in 
the complex plane for $\theta=0\,$ and different values of $q\,$. 
The diagonal phase shifts are evolved according to 
\begin{eqnarray}
\chi_{T0} &=& \chi_{T1}=t\;,
\nonumber\\
\chi_{T2} &=& -2\,t\;. 
\label{evol2qutrit1}
\end{eqnarray}
For the maximally entangled state ($q=0$), the overlap presents sharp 
peaks, touching the unit circle at the fractional phases expected 
for cyclic evolutions of qutrits. As $q$ is increased, the path followed 
in the complex plane degenerates to a circle for $q=1$ (product state). 

\begin{figure}[h!]
\includegraphics[scale=0.31]{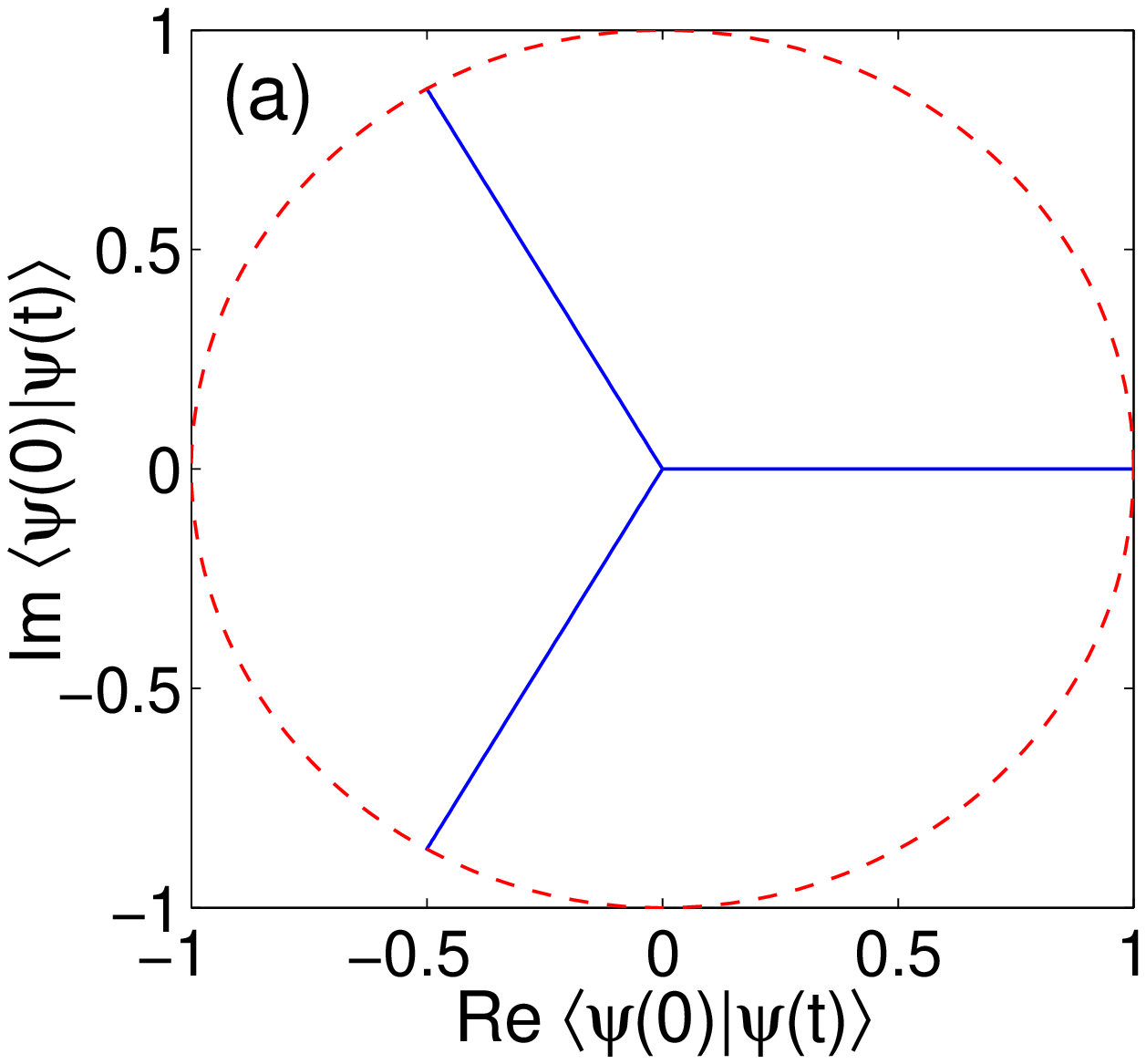}
\includegraphics[scale=0.31]{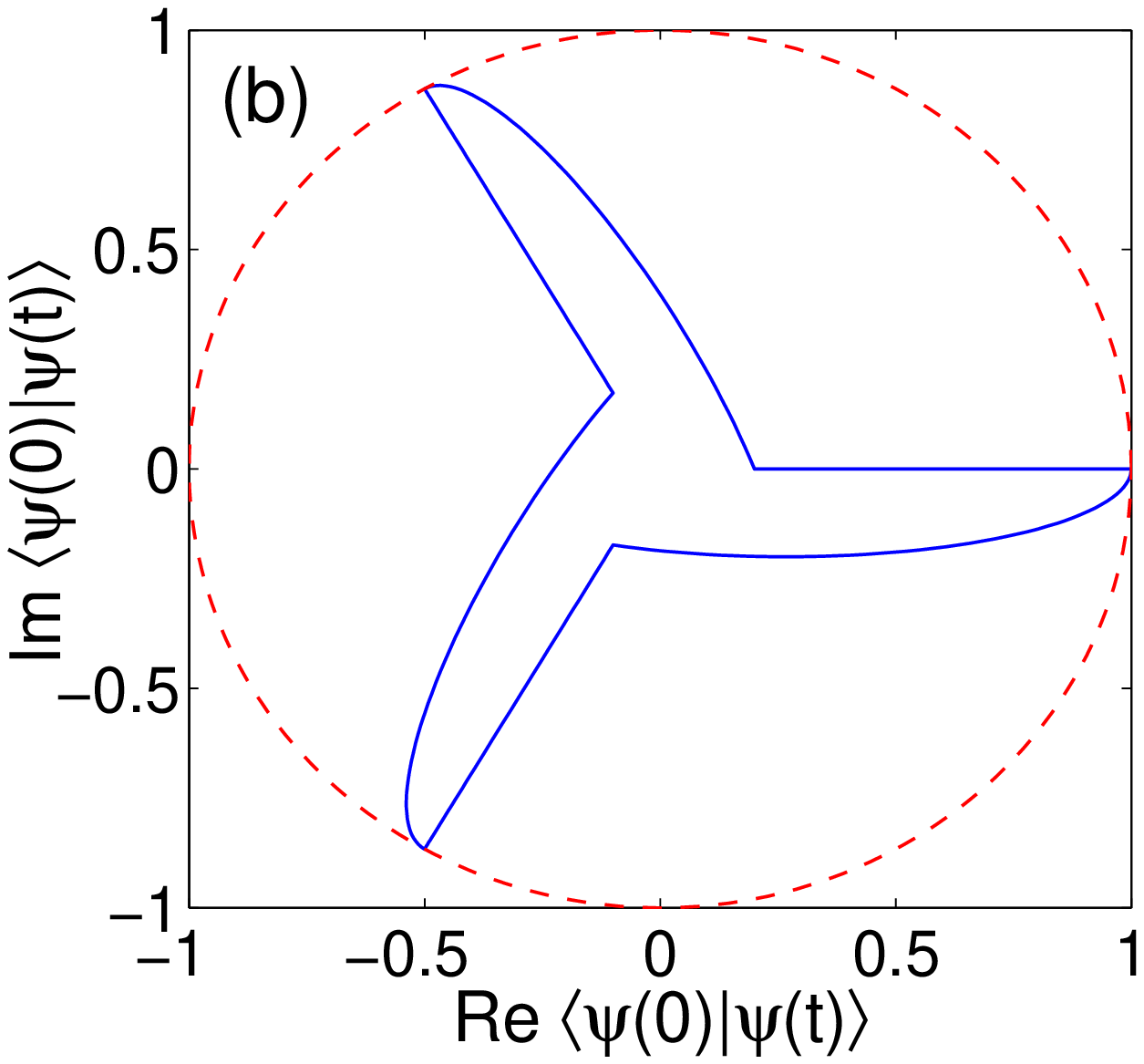}
\includegraphics[scale=0.31]{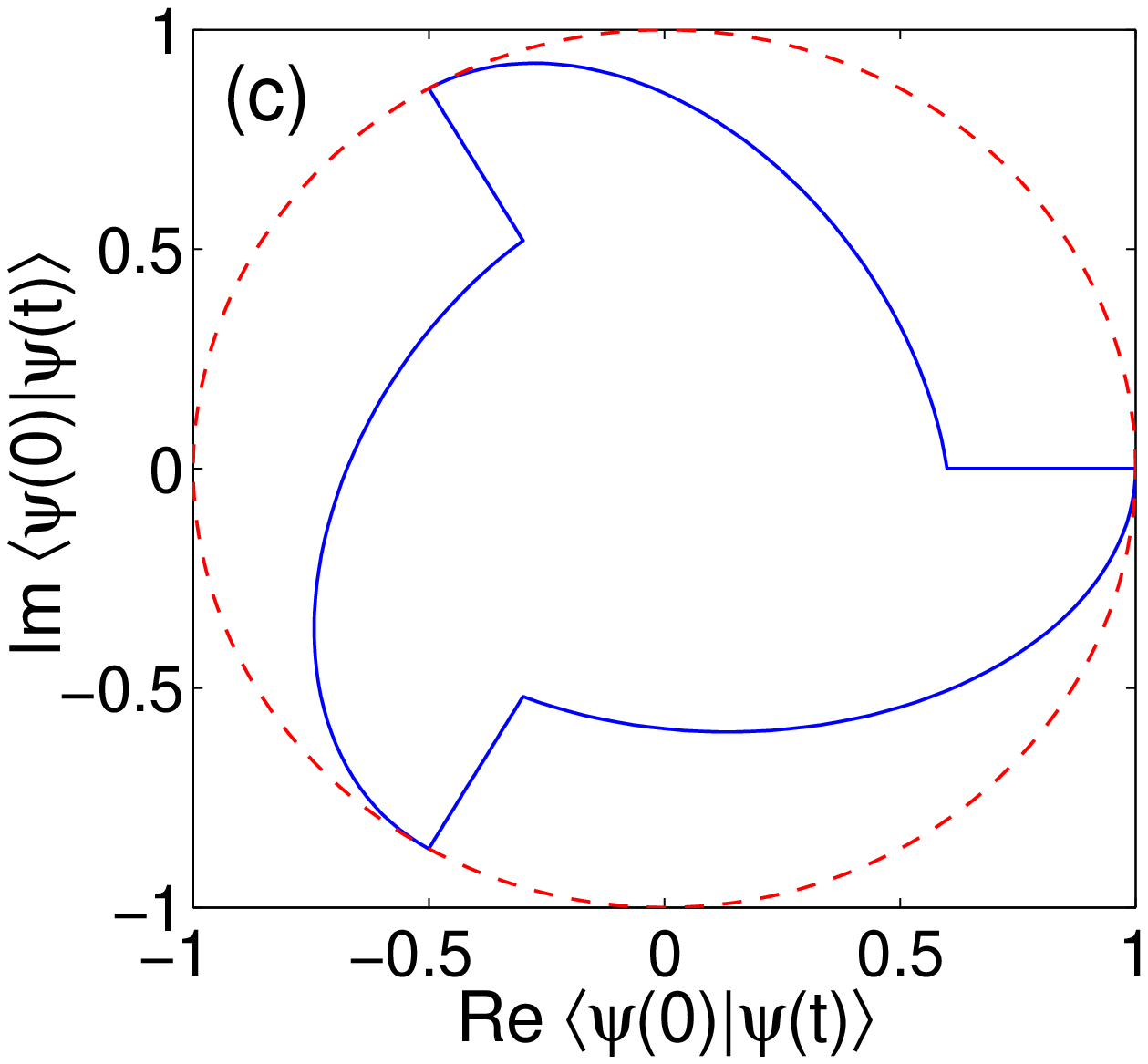}
\includegraphics[scale=0.31]{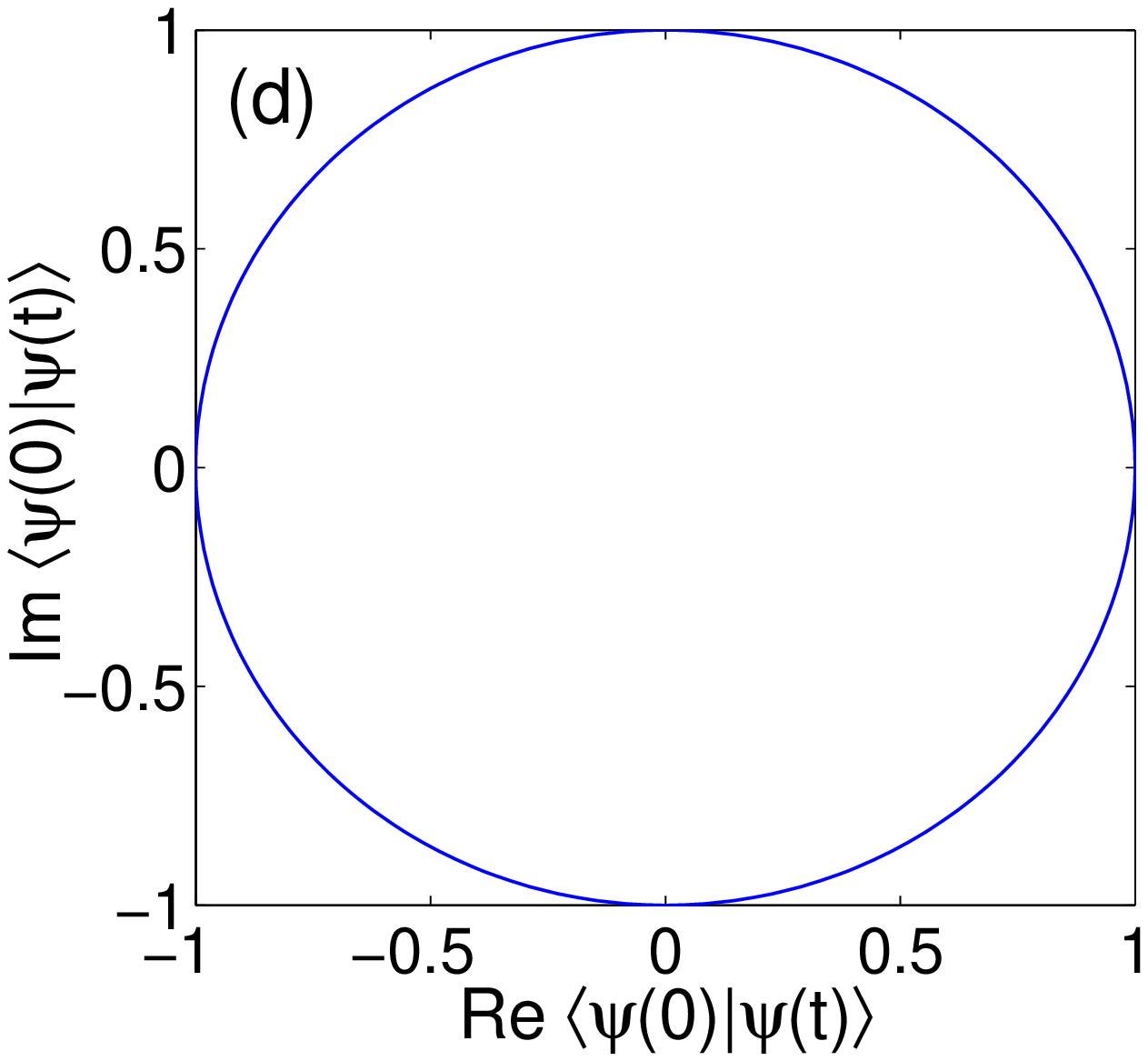}
\caption{\label{fig:sim2} Parametric plot of the quantum state overlap for a two-qutrit 
evolution given by Eqs.(\ref{evol2qutrit2}). 
(a) $q=0$ ($C=\sqrt{4/3}$), (b) $q=0.2\,$, (c) $q=0.6\,$, (d) $q=1$ ($C=0$). 
The unit circle is depicted in dashed red (online) for reference.}
\end{figure}

A second kind of evolution is considered in Fig.(\ref{fig:sim2}) in which 
the maximally entangled state follows sharp phase jumps between the 
fractional values. As entanglement is decreased, these jumps also 
degenerate to a continuous phase evolution (circle) for the product state. 
The diagonal phase shifts are evolved as 
\begin{eqnarray}
\chi_{T0} &=& -t\;, 
\nonumber\\
\chi_{T1} &=& \left\{
\begin{array}{ll}
t & (0\leq t\leq 2\pi/3)\\
2\pi/3 & (2\pi/3\leq t\leq 4\pi/3)\\
t-2\pi/3 & (4\pi/3\leq t\leq 2\pi)\\
4\pi/3 & (2\pi\leq t\leq 8\pi/3)\\
t-4\pi/3 & (8\pi/3\leq t\leq 10\pi/3)\\
2\pi & (10\pi/3\leq t\leq 4\pi)\;,
\end{array}\right.
\nonumber\\
\chi_{T2} &=& -(\chi_{T0}+\chi_{T1})\;. 
\label{evol2qutrit2}
\end{eqnarray}

\begin{figure}[h!]
\includegraphics[scale=0.45]{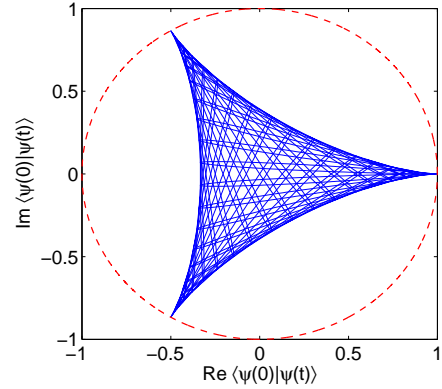}
\caption{\label{fig:sim3} 
Parametric plot of the quantum state overlap for a pair of 
maximally entangled qutrits under the joint 
evolution given by Eqs.(\ref{evol2qutrit3}). 
The unit circle is depicted in dashed red (online) for reference.}
\end{figure}

It is interesting to inspect how the overlap path is affected 
when the diagonal phase shifts evolve at very different speeds. 
For example, consider a maximally entangled state evolving according to 
\begin{eqnarray}
\chi_{T0} &=& t\;,
\nonumber\\
\chi_{T1} &=& 30\,t\;,
\nonumber\\
\chi_{T2} &=& -31\,t\;. 
\label{evol2qutrit3}
\end{eqnarray}
The corresponding result is displayed in Fig.(\ref{fig:sim3}). 
A complicated trajectory appears within the limits of the perimeter 
defined by Fig.(\ref{fig:sim1}a). This result raises the question 
about whether this fractional phase structure could still be observed 
under random local SU($3$) transformations. 

\begin{figure}[h!]
\includegraphics[scale=0.31]{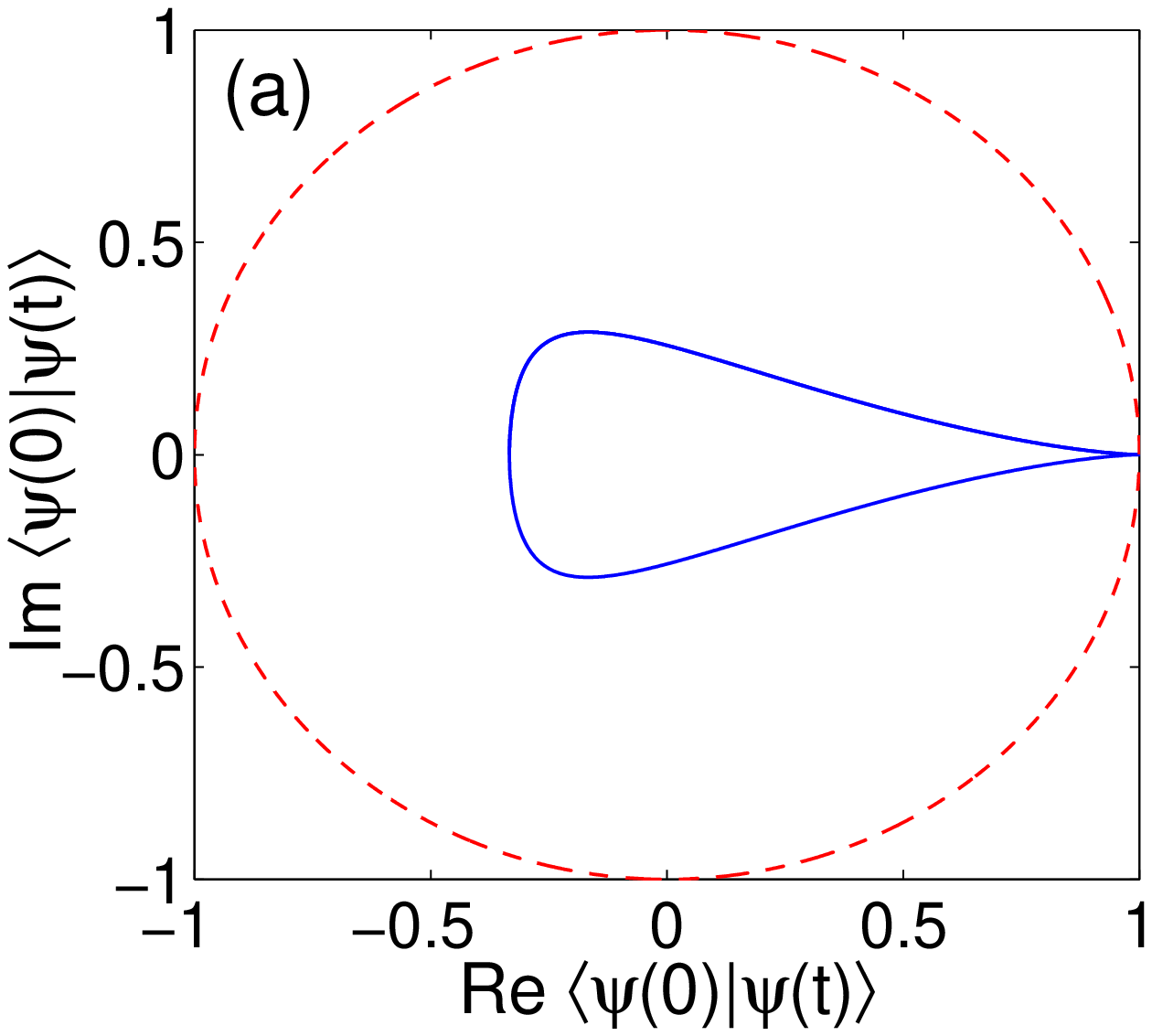}
\includegraphics[scale=0.31]{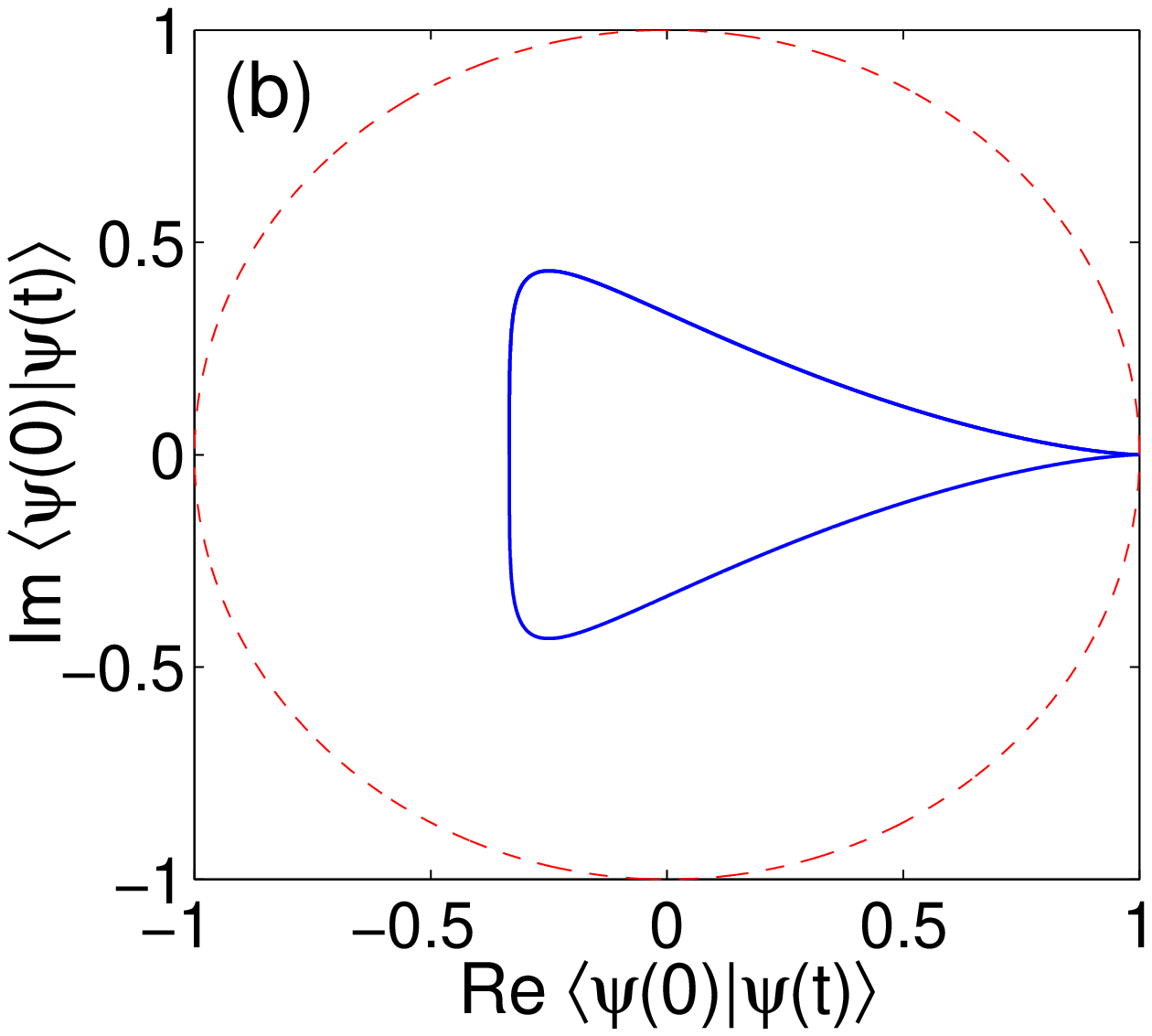}
\includegraphics[scale=0.31]{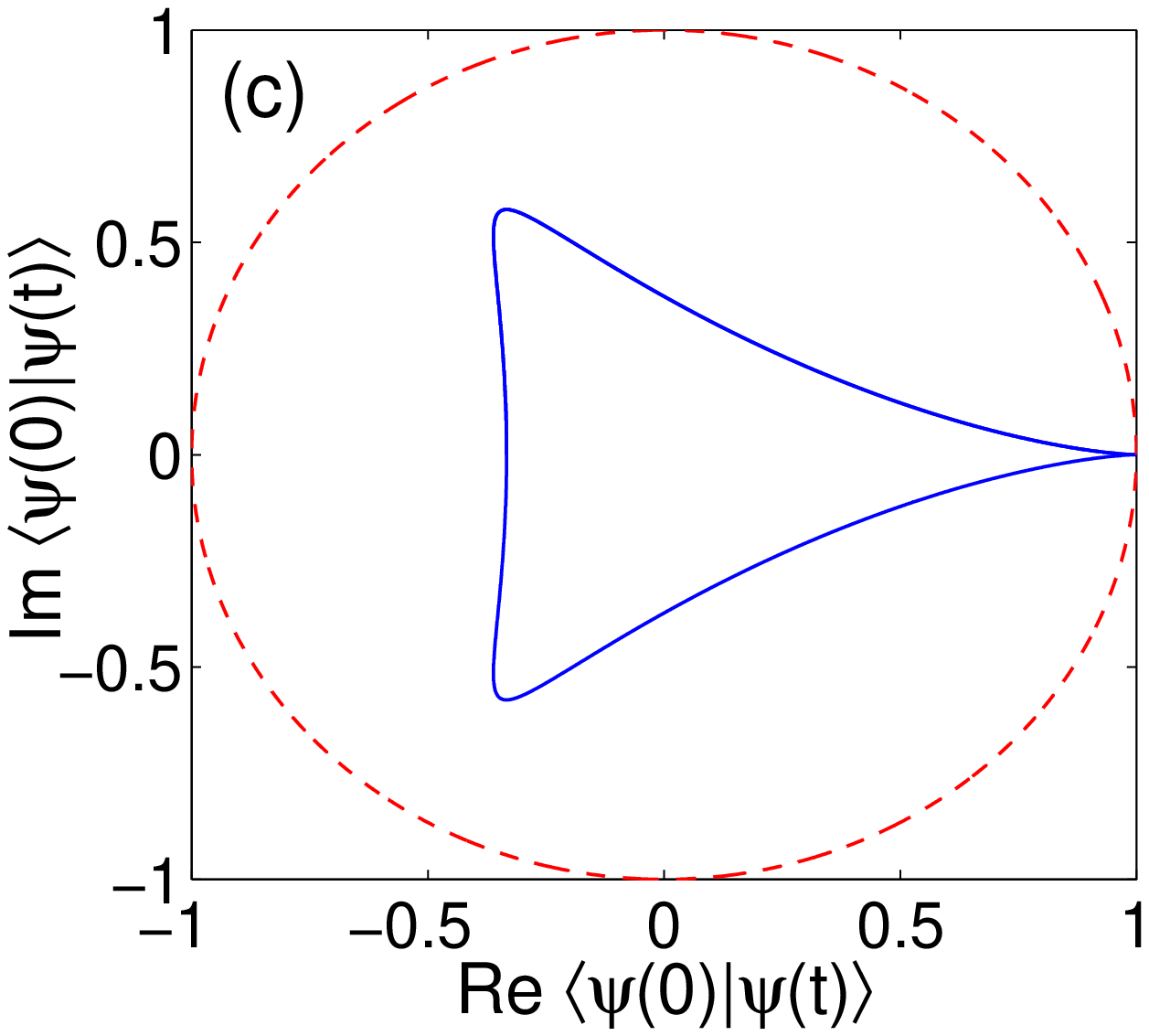}
\includegraphics[scale=0.31]{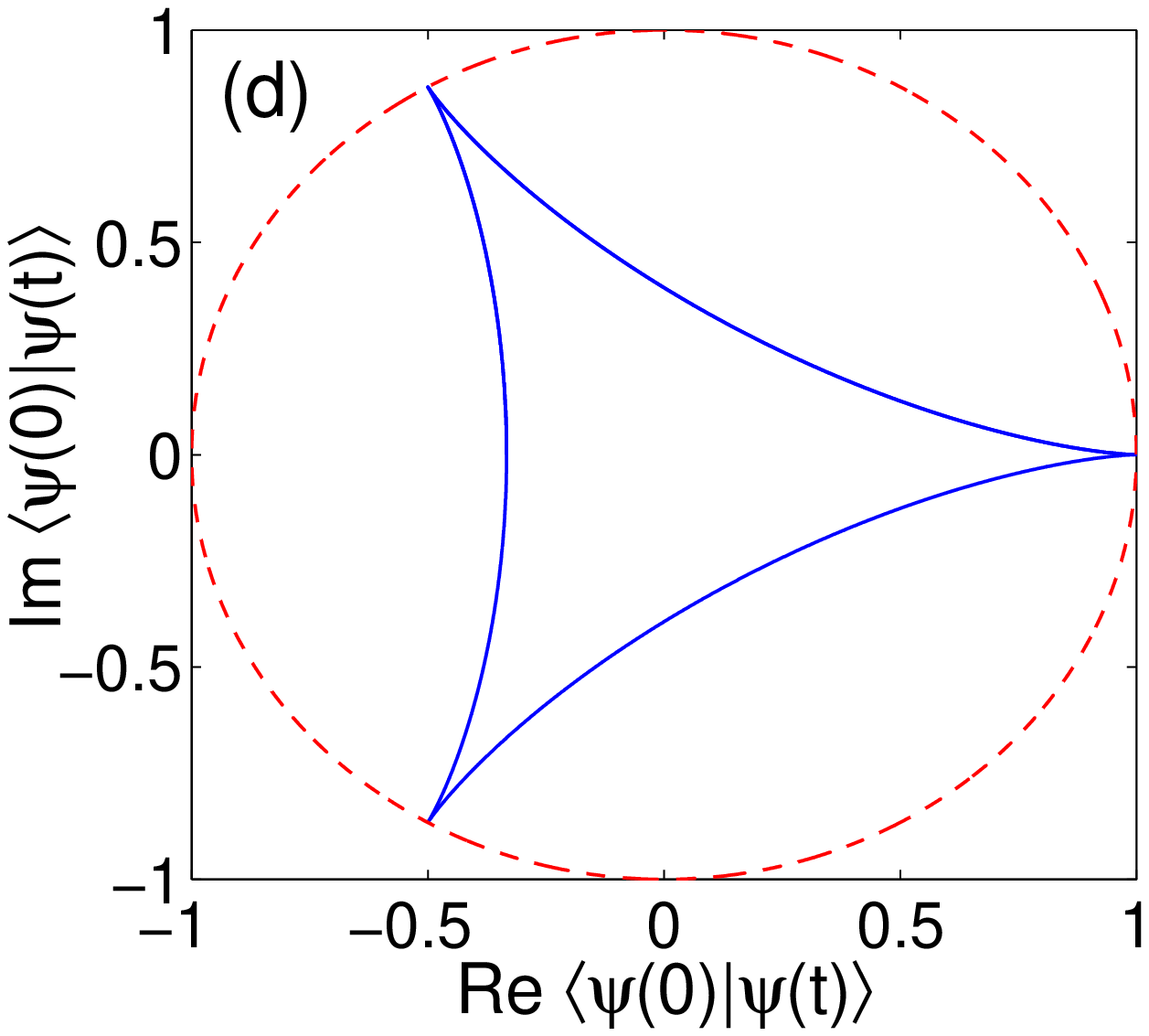}
\caption{\label{fig:sim6} Parametric plot of the quantum state overlap for a two-qutrit 
evolution given by Eqs.(\ref{evol2qutrit4}). 
(a) $q=1/3$ ($C=0$), (b) $q=1/2\,$, (c) $q=2/3\,$, (d) $q=1$ ($C=\sqrt{4/3}$). 
The unit circle is depicted in dashed red (online) for reference.}
\end{figure}

It is also interesting to compare the evolution of a maximally 
entangled state with partially entangled states having the 
same single qutrit probability distribution. For example, let 
us take the initial state 
\begin{eqnarray}
\ket{\psi(0)} &=& \sqrt{\frac{q}{3}}\,
\left(\frac{}{}\ket{00}+\ket{11}+\ket{22}\,\right) 
\nonumber\\
&+& \sqrt{\frac{1-q}{6}}\,\left(\frac{}{}\ket{01}+\ket{02}+\ket{12}\right.
\nonumber\\
&+& \left.\ket{20}+\ket{21}+\ket{10}\frac{}{}\right)\;, 
\label{2qutritstate2}
\end{eqnarray}
with $q$ ranging between $1/3$ for the product state and $1$ 
for the maximally entangled state. 
The probability distribution for qutrit $A$ is 
\begin{eqnarray}
P^A_n=\sum_{m=0}^{2} \abs{\braket{n\,m}{\psi(0)}}^2 =\frac{1}{3}\;, 
\end{eqnarray}  
for $n=0,1,2\,$, and similarly for $P^B_n\,$. To illustrate the 
role of entanglement, we can take the following parametric 
evolution
\begin{eqnarray}
\chi_{A0} &=& \chi_{A1} = t\;,
\nonumber\\
\chi_{A2} &=& -2\,t\;,
\nonumber\\
\chi_{B0} &=& \chi_{B1} = 2\,t\;,
\nonumber\\
\chi_{B2} &=& -4\,t\;, 
\label{evol2qutrit4}
\end{eqnarray}
where the local phase shifts are assymetrical. 
The parametric plot 
of the state overlap when the qutrits are subjected to the 
evolution given by Eqs.(\ref{evol2qutrit4}) is presented in 
Fig.(\ref{fig:sim6}). As we 
can see, only the maximally entangled state achieves maximum 
overlap at the fractional phases expected for qutrits. This 
comparison was used in refs.\cite{fracuffufmg} and \cite{multiqubit2} 
in the context of entangled photon pairs, where the quantum 
state overlap was associated to the visibility of two-photon 
interference fringes. Maximum visibility 
(overlap) can only occur with maximally entangled states 
at the allowed fractional phases. 

%

%%%%%%%%%%%%%%%%%%%%%%%%%%%%%%%%%%%%%%%%%%%%%%%%%%%%%%%%%%%%%%%%%%%%%%%%%%%%%%%%%%%%%%%%%%%%%%%

\subsection{Qubit-Qutrit}

We now illustrate the simplest case with assymetrical Hilbert spaces. 
Let us consider a qubit-qutrit system ($d_A=2, d_B=3$) initially 
prepared in the state
\begin{eqnarray}\ket{\psi(0)}=\frac{\sqrt{1+q\,}\,\ket{00}\,+\,
\sqrt{1-q\,}\,\ket{11}}{\sqrt{2}}\;,
\nonumber\\
\label{psiqubitqutrit}
\end{eqnarray}
with $0\leq q\leq 1\,$. The reduced density matrices of the 
qubit and the qutrit are
\begin{eqnarray}
\rho_A(0)&=& Q^2\;,
\nonumber\\
\rho_B(0)&=&\left[
\begin{matrix}
Q^2 & 0 \\
0   & 0 \\
\end{matrix}
\right]\;,
\end{eqnarray}
where
\begin{equation}
Q^2 = \frac{\mathbb{1}}{2}+\frac{\sqrt{1-C^2}}{2}\sigma_z\;,
\label{Q2qubitqutrit}
\end{equation}
and $C=\sqrt{1-q^2}$ is the qubit-qutrit concurrence. First, 
let us supppose that both the qubit and the qutrit evolve 
under local diagonal operations such that 
\begin{eqnarray}
\bar{U}_A(t)&=&\left[
\begin{matrix}
e^{i\,\chi_{A}} & 0 \\
0 & e^{-i\,\chi_{A}} \\
\end{matrix}
\right]\;,
\label{ubarABqubitqutrit}
\\
\bar{U}_B(t)&=& \left[
\begin{matrix}
e^{i\,\chi_{B0}} & 0 & 0 \\
0 & e^{i\,\chi_{B1}} & 0 \\
0 & 0 & e^{i\,\chi_{B2}} \\
\end{matrix}
\right]\;,
\nonumber
\end{eqnarray}
with $\chi_{B0}+\chi_{B1}+\chi_{B2}=0\,$. 
Then, the geometric phase acquired is 
\begin{eqnarray}
\phi_g &=& \arctan\left[\sqrt{1-C^2}\tan\left(\chi_A+\frac{\chi_{B0}-\chi_{B1}}{2}\right)\right]
\nonumber\\
&-&\sqrt{1-C^2}\,\left(\chi_A+\frac{\chi_{B0}-\chi_{B1}}{2}\right)\;.
\label{phigqubitqutrit}
\end{eqnarray}
This result is identical to the two-qubit geometric phase if we make the 
identification $\chi_B\equiv(\chi_{B0}-\chi_{B1})/2\,$. 
Note that state (\ref{psiqubitqutrit}) 
does not include the third component of the qutrit, which remains 
unaffected as long as only diagonal operations are performed. 
Thus, the qutrit behaves as an effective qubit and only two-qubit 
fractional phases can be observed. 

In order to evidence the dual dimensional structure of the qubit-qutrit 
system, using only diagonal evolutions, we must consider an initial state 
with all qubit and qutrit components. We can build a simple numerical 
example with the following maximally entangled state
\begin{eqnarray}
\ket{\psi(0)}=\frac{1}{\sqrt{2}}\,\ket{00}\,+\,
\frac{1}{2}\,\ket{11}\,+\,\frac{1}{2}\,\ket{12}\;, 
\label{psiqubitqutrit2}
\end{eqnarray}
for which $Q^2=\mathbb{1}_{2\times 2}/2\,$, 
$S_A(0)=\mathbb{1}_{2\times 2}\,$, and 
\begin{eqnarray}
S_B(0)&=&\left[
\begin{matrix}
1 & 0 & 0 \\
0 & \;\;\frac{1}{\sqrt{2}} & \frac{1}{\sqrt{2}} \\
0 & -\frac{1}{\sqrt{2}} & \frac{1}{\sqrt{2}} \\
\end{matrix}
\right]\;.
\end{eqnarray}
According to Eq.(\ref{phigfrac}), the integral contribution for the geometric phase 
never vanishes for qudits with different dimensions, even for maximally entangled states. 
The dual dimension behavior, however, will still be present in the nontrivial total 
phase $\bar{\phi}_{\,tot}\,$. 
Let us consider the local diagonal evolutions given by Eqs. (\ref{ubarABqubitqutrit}). 
In this case, we obtain
\begin{eqnarray}
\bar{\phi}_{\,tot}&=&\arg\left\{\cos\left(\chi_A-\chi_{B2}-\frac{\chi_{B1}}{2}\right)\,
\frac{e^{-i\,\chi_{B1}/2}}{2}
\right.\nonumber\\
&+&\left.\cos\left(\chi_A-\chi_{B1}-\frac{\chi_{B2}}{2}\right)\,
\frac{e^{-i\,\chi_{B2}/2}}{2}\right\}\;,
\end{eqnarray}
and 
\begin{equation}
\phi_g=\bar{\phi}_{\,tot}-\frac{\chi_{B0}}{4}\;. 
\end{equation}

\begin{figure}[h!]
\includegraphics[scale=0.31]{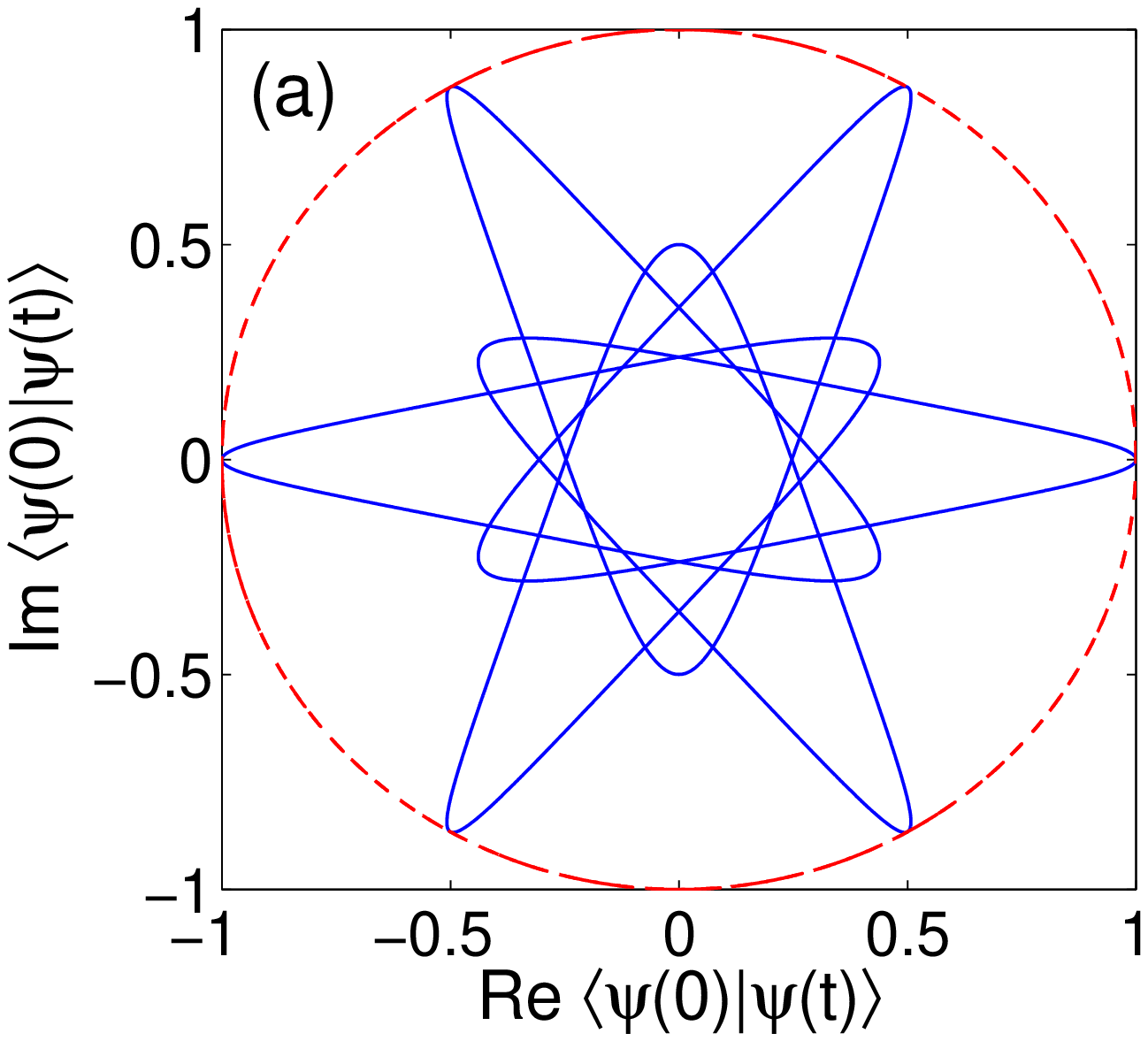}
\includegraphics[scale=0.31]{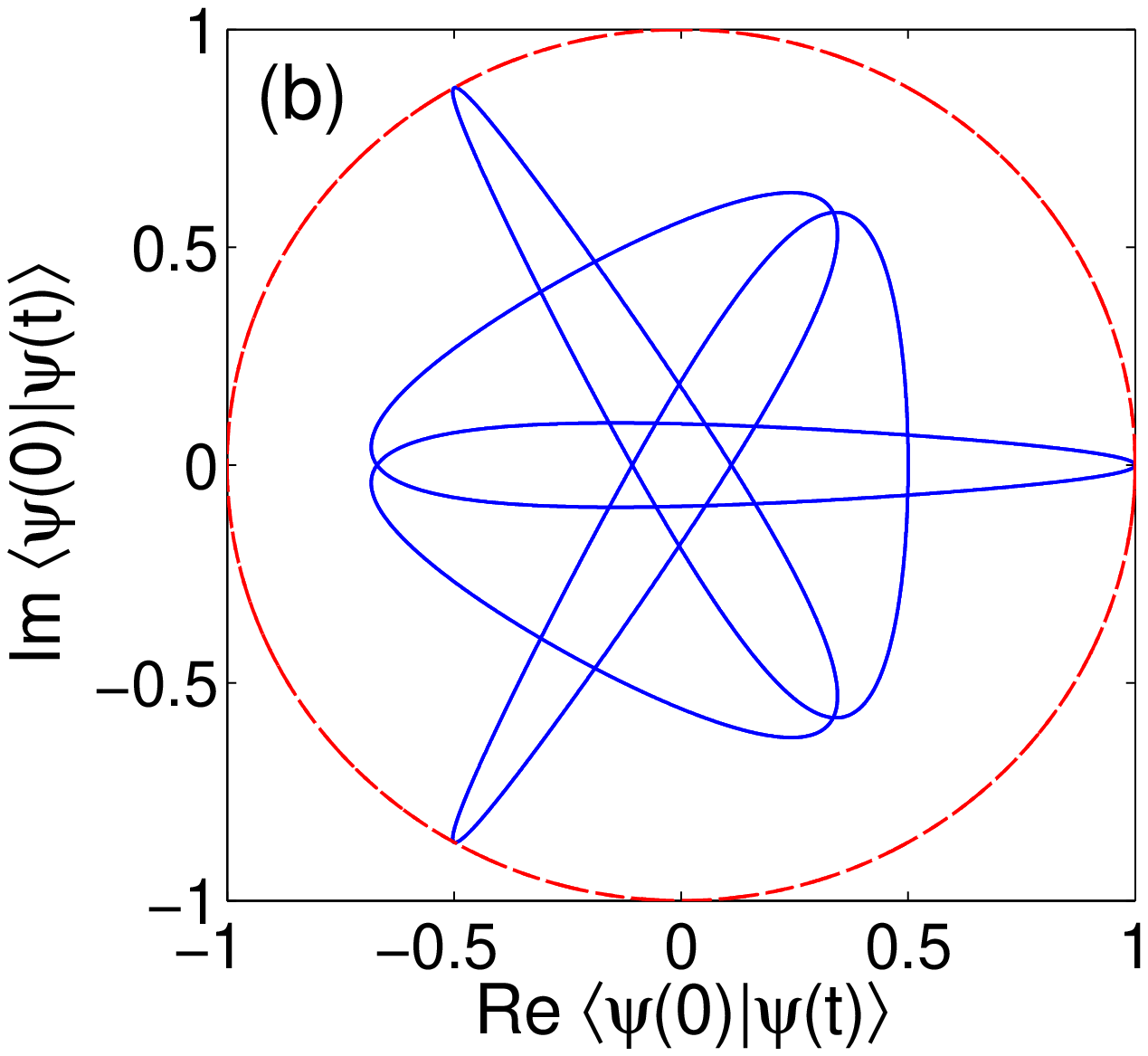}
\includegraphics[scale=0.31]{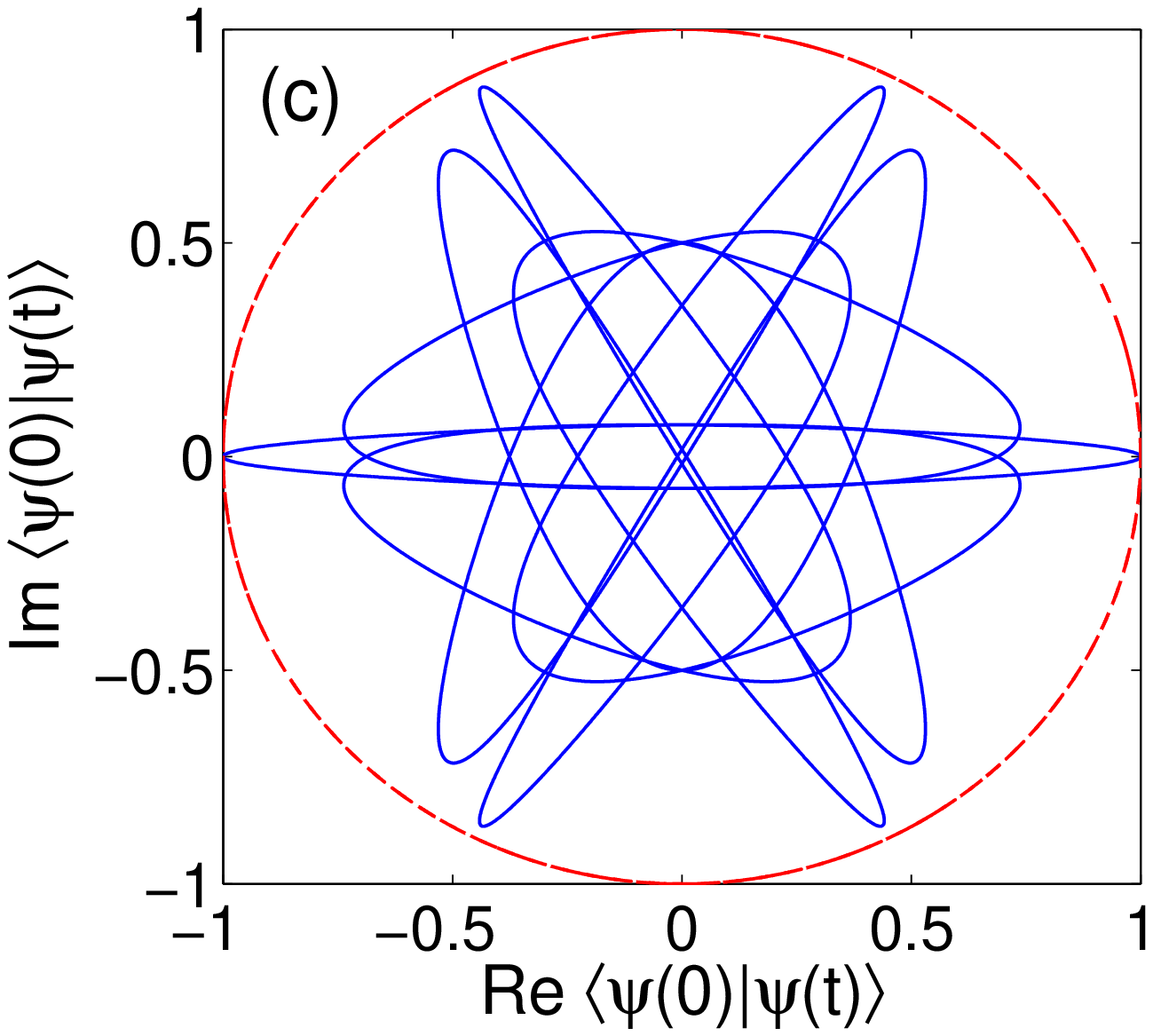}
\includegraphics[scale=0.31]{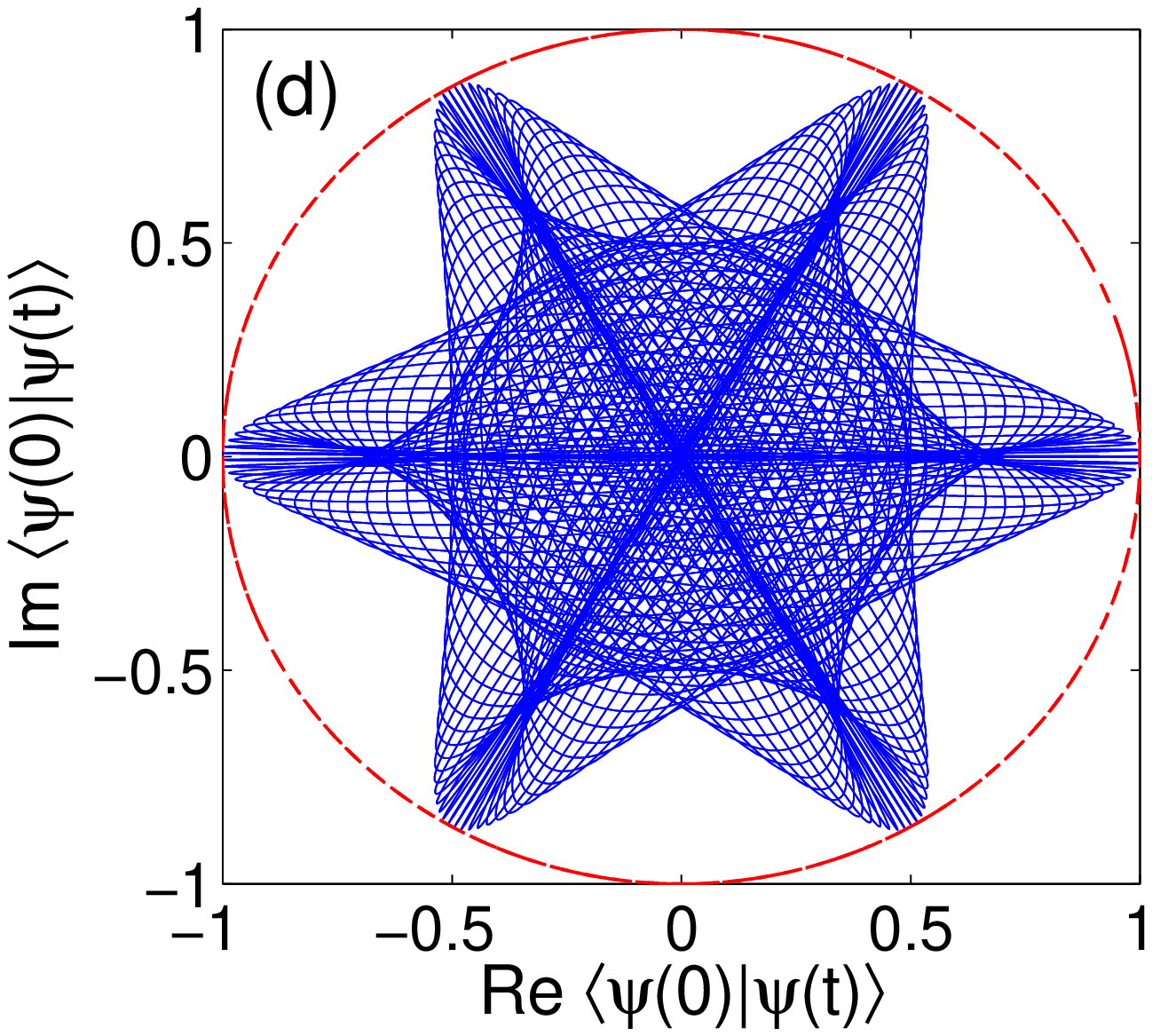}
\caption{\label{fig:sim4} Parametric plot of the qubit-qutrit quantum 
state overlap for the initial condition given by Eq.(\ref{psiqubitqutrit2}). 
(a) $\chi_{A}=1.5\,t\,$, (b) $\chi_{A}=3\,t\,$, (c) $\chi_{A}=3.5\,t\,$ 
(d) $\chi_{A}=100\,t\,$. 
The unit circle is depicted in dashed red (online) for reference.}
\end{figure}

The fractional phases expected for a qubit-qutrit system are 
\begin{eqnarray}
\bar{\phi}_{tot} = n\pi + \frac{2\,m\,\pi}{3}\;, 
\end{eqnarray}
with $n,m\in\mathbb{Z}\,$. Therefore, the qubit-qutrit system 
can exhibit the qubit, the qutrit or a combination of both 
topological phases when subjected to cyclic evolutions under 
SU$(2)\,\otimes$ SU($3$) operations. In Fig.(\ref{fig:sim4}) 
different diagonal evolutions are considered. In all cases 
the qutrit is operated by 
\begin{eqnarray}
\chi_{B0} &=& \chi_{B1} = t\;,
\nonumber\\
\chi_{B2} &=& -2\,t \;, 
\label{evolBqubitqutrit}
\end{eqnarray}
while different evolutions are considered for the qubit. 
The dual phase behavior can be observed in Fig.(\ref{fig:sim4}a), while 
Figs.(\ref{fig:sim4}b) and (\ref{fig:sim4}c) display the qutrit and 
qubit phases respectively. 
The dual phase behavior also becomes evident when we make the qubit 
evolution much faster than the qutrit, as in Fig.(\ref{fig:sim4}d). 
In this case, a complicated path is drawn by the 
quantum state overlap in the complex plane, touching the 
unit circle only when the fractional phases allowed for 
the qubit-qutrit system are attained. 

%%%%%%%%%%%%%%%%%%%%%%%%%%%%%%%%%%%%%%%%%%%%%%%%%%%%%%%%%%%%%%%%%%%%%%%%%%%%%%%%%%%%%%%%%%%%%%%

\section{conclusion}
\label{conclusion}

In this article, we presented a detailed description of the geometric phase 
acquired by entangled qudits operated by local unitary transformations. 
Our previous result \cite{fracuff} was detailed and extended to pairs of qudits 
with general dimensions $d_A$ and $d_B$. This was achieved by utilizing the singular 
value decomposition of the coefficient matrix defined by the two-qudit 
quantum state. This decomposition involves a pair of matrices in SU($d_A$) and
SU($d_B$), respectively, with the dimensions of the individual qudit Hilbert spaces. The 
fractional phase values naturally appear as the possible factors 
arising from cyclic evolutions of these local components. They completely encompass the geometric 
phase acquired by maximally entangled qudits with equal dimensions, subjected to cyclic 
evolutions. However, in a more general scenario where partially 
entangled states or different qudit dimensions are considered, 
the geometric phase can assume continuous values in addition 
to the fractional phase contribution. 

To put in evidence the role played by entanglement and the SU($d$) parameters 
of the local transformations, we used the geometric phase derived by Mukunda and Simon 
in Refs.\cite{smukunda,smukunda2}, as well as the decomposition of the local transformations 
into the Cartan and coset ${\rm SU}(d)/{\rm U}(1)^{d-1}$ sectors. In particular, 
we showed that the geometric phase given by Mukunda and Simon gives rise to a 
holonomic contribution built in the Cartan sector and a 
nonholonomic one built in the coset sector. 
Our results regarding the fractional phases in higher dimensions were illustrated 
with numerical examples for two-qutrit 
and qubit-qutrit systems. This investigation could be applied to different experimental contexts, 
including entangled photon pairs created by spontaneous parametric 
down conversion, nuclear magnetic resonance, trapped ions, and 
other setups dealing with entangled states.

Qudit gates based on topological phases are a potentially robust means 
to implement quantum algorithms \cite{bullock,ashok,munro}. 
In order to demonstrate the usefulness of the fractional phases for 
quantum information protocols, it will be crucial to investigate 
the phase evolution under local random unitary transformations. 
It is well known that two-qubit entangled states are robust 
against certain kinds of noise \cite{cryptosteve}, what motivated 
an alignment free quantum cryptography protocol \cite{cryptouff,cryptolorenzo}. 
We shall leave the investigation of the fractional phases under noisy 
evolutions to a future contribution.

\section*{Acknowledgments}
We are grateful to E. Sj\"oqvist and M. Johansson for useful 
discussions. 
Funding was provided by 
Conselho Nacional de Desenvolvimento Tecnol\'ogico (CNPq), 
Coordena\c c\~{a}o de Aperfei\c coamento de 
Pessoal de N\'\i vel Superior (CAPES), Funda\c c\~{a}o de Amparo \`{a} 
Pesquisa do Estado do Rio de Janeiro (FAPERJ-BR), and Instituto Nacional 
de Ci\^encia e Tecnologia de Informa\c c\~ao Qu\^antica (INCT-CNPq).

\end{document}